# The Link Between Large Scientific Collaboration and Productivity. Rethinking How to Estimate the Monetary Value of Publications[1],[2]

Francesco Giffoni[a]*, Louis Colnot[a], Emanuela Sirtori[a]

[a]CSIL, Corso Monforte 15, 20122, Milan, Italy.

Pre-print. Version draft: 13.11.2024

______________________________________________________________________

**Abstract**

This paper addresses how to assign a monetary value to scientific publications, particularly in the case of multi-author papers arising from large-scale research collaborations. Contemporary science increasingly relies on extensive and varied collaborations to tackle global challenges in fields such as life sciences, climate science, energy, high-energy physics, astronomy, and many others. We argue that existing literature fails to address the collaborative nature of research by overlooking the relationship between co-authorship and scientists' productivity. Using the Marginal Cost of Production (MCP) approach, we first highlight the methodological limitations of ignoring this relationship, then propose a generalised MCP model to value co-authorship. As a case study, we examine High-Energy Physics (HEP) collaborations at the Large Hadron Collider (LHC) at CERN, analysing approximately half a million scientific outputs by over 50,000 authors from 1990 to 2021. Our findings indicate that collaborative adjustments yield monetary valuations for subsets of highly collaborative papers up to 3 orders of magnitude higher than previous estimates, with elevated values correlating with high research quality. This study contributes to the literature on research output evaluation, addressing debates in science policy around assessing research performance and impact. Our methodology is applicable to authorship valuation both within academia and in large-scale scientific collaborations, fitting diverse research impact assessment frameworks or as self-standing procedure. Additionally, we discuss the conditions under which this method may complement survey-based approaches.

______________________________________________________________________

*Corresponding author

Email addresses: giffoni@csilmilano.com (Francesco Giffoni), colnot@csilmilano.com (Louis Colnot), sirtori@csilmilano.com (Emanuela Sirtori).

___________________________

[1] **Funding.** The research presented in this paper was funded by the Future Circular Collider Innovation Study (FCCIS), a project supported by the European Union's Horizon 2020 research and innovation programme under grant No. 951754.

[2] **Acknowledgements**: We wish to thank Christian Caron (Springer Nature), Johannes Gutleber (CERN), and Valentina Morretta (University of Milan) for their help to enrich and strengthen the content of this paper. We are also grateful to Stefano Carrazza (University of Milan and CERN) and Massimo Florio (CSIL) for their comments on an earlier draft of the paper. Thank you to Davide Vurchio (University of Bari) and Erica Delugas (CSIL) for sharing their salary data.

# 1. Introduction

Contemporary science is distinguished by expanding and diverse forms of collaboration that span organizational, disciplinary, cultural, and national boundaries (Qi et al., 2023; Hall et al., 2018).[3] In STEM fields (science, technology, engineering, and mathematics), over 90% of research studies and publications involve collaboration (Bozeman and Boardman, 2014). At the global level, the share of publications from international collaboration increased from 4.7% in 1980 to 25.7% in 2021 (Aksnes and Sivertsen, 2023). There are many reasons why research collaboration is increasingly expanding. Today's most pressing issues, such as climate change, life science, energy and sustainable development, are global in nature, and collaborative research is essential for developing new scientific knowledge and comprehensive strategies to tackle them (Abbass et al., 2022; Jessel et al., 2019).

However, the meaning of international collaboration extends beyond the concept of setting teams of individuals to achieve the common goal of producing new knowledge (Bozeman et al., 2014; Katz and Martin, 1997). Research funds, including European programmes such as Horizon Europe, request more interdisciplinary and collaborative research to bridge the gap between research and practice (Dwivedi et al., 2024; Sibbald et al., 2019). Collaborative efforts allow for resource sharing, equipment, software and data, which can be particularly valuable for expensive or resource-intensive projects that isolated teams and countries cannot afford alone (Wagner et al., 2015), such as large-scale research infrastructures (ESFRI, 2021).[4] Supra-national coordination, collaboration and joint investment in world-leading research and technology infrastructures is essential not only for ground-breaking R&I, but also for the sustainability and competitiveness of geographical ecosystems, such as the European ones (Draghi Report, 2024). Research competition embeds universities as both individual and collective actors simultaneously in diverse nested and interdependent networks, where the cross-pollination of ideas is expected to lead to breakthroughs that might not occur in isolated settings (Krücken 2019; Musselin, 2018).

Despite the importance of collaboration for scientific research, the economic literature struggles to quantify its value accurately. A recent stream of literature deals with the estimation of the monetary value of scientific publications, but it largely overlooks the role of collaboration, proxied by co-authorship (Katz and Martin, 1997), in contributing to that value (Morretta et al., 2022, Bastianin et al., 2022; Rosseau et al., 2021, Florio, 2019; Bastianin and Florio 2018; Florio et al., 2016; Florio and Sirtori, 2016; Battistoni et al., 2016). This literature either treats scientific papers as "single-author" products or gives equal credit to all coauthors. It does not account for how collaboration influences the value of publications (Florio ad Sirtori, 2016). The consequence is a biased, mainly underestimation of the publications' value. Accurately estimating the monetary value of scientific publications is important for several reasons. Not only does it help assess research productivity, but it also expands our understanding of the broader impacts of research. This, in turn, supports informed decision-making in research policy. Furthermore, assigning a monetary value to publications highlights the economic benefits of scientific research, which frequently drives technological advancements, innovations, and economic growth. Therefore, addressing the collaborative element is crucial for providing a more comprehensive and accurate estimation of research value, ensuring that the broader benefits of scientific collaboration are recognised.

---

[3] In this paper, research (international) collaboration is defined as an organisation leading to publications having authors affiliated with institutions located in the same (or more than one) country (Katz and Martin, 1997).
[4] We use big-science projects and large -scale research infrastructures interchangeably.

We attain two main results.

1. Integrating collaboration-related metrics into our evaluation framework, our findings point to a monetary value of a subset of highly collaborative papers up to 3 orders of magnitude higher than previously reported. The value depends critically on the accurate measurement of the authors' productivity function.
2. In line with the predictions of the literature on the relationship between co-authorship and scientific excellence (Thelwall et al., 2023; Isfandyari-Moghaddam et al., 2023; Abbasi et al., 2011), we show that a greater number of co-authors correlates with a higher monetary value of the scientific outputs.

We contribute to the above stream of literature from at least three standpoints.

First, we address an under-researched area in the economic evaluation of science by monetarily evaluating scientific publications using free, accessible data to anyone. While estimating the value of scientific products is a pressing need in research policy, there are still many open issues, and further methodological and empirical research is necessary (Rosseau et al., 2021).

Second, we innovate methodologically by generalising existing practices to accommodate multi-author papers resulting from research collaboration. The topic is highly relevant for scientific fields where co-authorship is the standard practice, such as physics, astronomy, climate and life science. To the best of our knowledge, literature often treats such scientific fields as outliers. For instance, Nielsen and Andersen (2021) exclude physics and astronomy from their analysis as the number of coauthors (>1,000) on some articles skewed the overall trends of all included fields. Similarly, Andersen (2023) treats papers with more than 100 authors as extreme cases to be removed. In Rosseau et al. (2021), only 59 answers (18% of the sample) related to mathematics, physics and natural sciences (see Table 3, p. 6 in their paper), making their analysis weak to say something on co-authorship in these fields. Florio (2019: 97) acknowledges that, when estimating the author's productivity, multiple authorship needs to be considered, but does not explain how this should be done. Actually, he suggests that *"[…] practical solution should be found case by case."*[5]

---

[5] Regarding productivity, Florio (2019:97) says: "in cases of multiple authorship, an issue to be solved when estimating the average productivity is the definition of the individual contribution to an article (e.g. life science has different style in listing the sequence of authors in comparison to other fields)". Attention should be paid to two situations. Experiment collaborations articles are typically signed by a long list of authors, including many scientists who did not directly contributing to writing the text. Conversely, in other cases, the article is signed only by those who wrote it, although may be that more contributed to the experimental work in certain role. In both cases the authors productivity estimate can be distorted if these aspects are not appropriately reflected in the calculation practical solution should be found case by case…"

Instead, we show how the method based on the marginal cost of production (hereafter MPC) can be generalised to account for collaboration. This approach can be applied rigorously and objectively across different scientific fields, minimising the number of arbitrary assumptions that would arise if solutions were implemented on a case-by-case basis. On top of that, and since other valid approaches exist to value scientific publications, we also discuss when the MPC can be combined with other approaches (e.g., survey-based methods) in the spirit of Rosseau et al. (2021).

Third, we perform a sensitivity analysis on the main parameters of our model to assess the impact of the assumptions on our results. Indeed, scientists' yearly productivity, the working time they devote to research, and the number of authors per paper differ from one discipline to another due to various practices in use (Abrahams et al. 2019; Florio, 2019; Flores-Szwagrzak and Treibich, 2015). We shed light on the relationship between collaboration in research and scientists' productivity, which in turn influences the monetary value of scientific publications. According to Rosseau et al. (2021:12), "*It would be interesting […] to study how publication characteristics such as a number of coauthors influence the estimation of the value of publications*". Similarly, Morretta et al. (2022) argue that the share of time dedicated to research can fit certain situations in some institutions or countries but may not necessarily be realistic in other contexts. Our paper addresses these open issues.

The scientific collaborations at the Large Hadron Collider (LHC) at CERN, the world's leading laboratory for experimental particle physics, serve as an exemplary testing ground for our study. Bringing scientists together across national boundaries to boost science and advance human knowledge about what the universe consists of has been one of the primary goals of CERN since its establishment in 1954.[6] At CERN, LHC is an avatar of international science collaboration. Since 1990, 34,558 LHC-related scientific products[7] (40% in collaboration between at least two authors) have been produced by 50,659 different authors, offering an enormous amount of data and a unique opportunity to reliably test our methodology.[8] On top of that, it has been used as a benchmark for this type of analysis, giving a chance to discuss our research's result vis-à-vis existing evidence (Bastianin et al., 2022; Florio 2019; Bastianin and Florio, 2018; Florio et al., 2016). Accordingly, CERN LHC provides an ideal case study for our research objective. The insights developed in this paper can be easily applied to different scientific environments from HEP, such as genomics, public health, and climate science, astronomy, to mention a few, where they provide new ways of measuring the monetary value of publications in case of collaboration.

The rest of the article proceeds as follows. Section 2 provides a literature review to introduce the conceptual model and our research hypotheses in Section 3. Section 4 briefly presents CERN and LHC's collaboration practices and organisation. Building on the context provided in Section 4, Section 5 outlines the data and the step-by-step statistical approaches used to process it, ensuring that the procedure is easily replicable across different scientific fields. The results are presented in Section 5. Section 6 concludes with a discussion of policy implications, caveats, and suggestions for future research, including a discussion of the MPC approach with alternative approaches.

---

[6] See https://home.cern/about/who-we-are/our-mission Last access in October, 2024.
[7] In this paper, we use the term scientific products to group: (i) research articles in peer-reviewed journals, pre-prints, conference proceeding and papers, doctoral theses, scientific notes, book and book chapters.
[8] These numbers refer to the period 1990- 2021, covering both the LHC construction phase (1993 – 2007) and the operating phase from 2008 to 2021. The number of authors is based on the ones that could be unambiguously identified in the INSPIRE HEP database (see Section 5) and therefore, it is likely a lower bound.

## 2. Literature review

### 2.1 Approaches to evaluate scientific products

There are three main evaluation approaches for inferring the value of scientific products, each with advantages and caveats. A snapshot of each is provided below (see Rosseau et al., 2021 for an excellent review).

The first approach looks at the prices for publications in the existing academic publishing market (Larivière et al., 2015; McGuigan, 2004; Chressantis and Chressantis, 1993). The oligopoly structure of the market with a few and big publishers makes prices distorted signals for economic valuation, not reflecting the true value of publications (Rosseau et al., 2021). On top of that, the increasing number of online journals and open-access initiatives reduces opportunities to collect information on the observed prices while not leading to a big shift in the market structure (Larivière et al., 2015). For instance, the field of High Energy Physics (HEP) has explored alternative communication strategies for publication in peer-reviewed journals for decades, initially via the mass mailing of copies of preliminary manuscripts, then via the inception of the first online repositories and digital libraries like (ArXiv, INSPIRE-HEP, CERN Document Server (CDS). Gentil-Beccot et al. (2009) show that free and immediate online dissemination of pre-prints creates an immense citation advantage in HEP, whereas publication in Open Access journals presents no discernible advantage.[9] The authors also show that HEP scientists rarely read journals, opting for pre-prints instead. Aside from incomplete pricing information, the key takeaway from Gentil-Beccot et al. (2010) for this study is that the scope of the analysis must be expanded to include a broader range of scientific outputs beyond peer-reviewed journal publications. These include pre-prints, technical reports, conference proceedings, papers, notes, theses, book chapters, and more. On the positive side, the market approach can be easily implemented, data are easily available, and the idea is understandable even to non-scientific audiences, including policymakers.

The second approach is the marginal production cost (MPC). It argues that the marginal social value of a publication is at least equal to the cost (including the salary of the authors) to produce it (Florio and Sirtori, 2016; European Commission, 2014). The MPC approach, as implemented so far, faces several methodological issues (outlined below), particularly given that the academic labour market is heavily subsidised (Morretta et al., 2022; Rosseau et al., 2021). Moreover, regulations and transaction costs are likely to limit the international mobility of researchers, resulting in observed salaries being distorted and away from a well-functioning market (Rosseau et al., 2021). That said, we argue that the marginal cost of production can still be a valid method for valuing publications, provided it accounts for diverse publication behaviours across scientific fields, including collaboration, while reducing arbitrary assumptions.

---

[9] Gentil-Beccot (2009) is pre-SCOAP3 (Sponsoring Consortium for Open Access Publishing in Particle Physics). SCOAP3 is a further channel to disseminate results to the global scientific community. Since 2014 this has changed considerably, e.g. access to journals had taken over again over Arxiv access for the SCOAP3 journals. See https://scoap3.org/. Last access in November, 2024.

The third approach relies on survey-based valuation techniques such as contingent evaluation and discrete choice experiments (Rosseau et al., 2021). They are valid alternatives to elicit the willingness-to-pay (WTP) for scientific publications, but they are vulnerable to hypothetical bias and non-consequentiality (Giffoni and Florio, 2023; Johnston et al., 2017). Unlike the MPC, which focuses on the supply side, these techniques value scientific publications from the demand side, considering the consumer's perspective. The goal is to assess how much scientists would be willing to pay to benefit from an additional publication in their field.

Surveys deal with statements instead of real behaviours, and respondents' choices cannot reflect real commitments, resulting in biased values of the WTP. Strict guidelines to minimise such risks exist (Johnston et al., 2017), making this kind of survey feasible but very costly and time-consuming, often deterring the implementation of the survey itself or applicable to limited contexts and targets that undermine external validity. That said, primary data could offer further insights into the estimation of the social value of publications, including non-market components such as externalities (Morretta et al., 2022) and non-use values (Rosseau et al., 2021).

## 2.2 Evaluating scientific output through the marginal cost of production (MPC): current state-of-the-art

The current applications of the MPC in the research market aim to capture the social value that an additional publication brings to the scientific community by examining its production cost. The MPC does not measure the inherent value of scientific knowledge contained in the paper, such as its contribution to a specific field, new insights, breakthrough discoveries, or the academic recognition it may bring to the authors (Li et al., 2013).[10] This is particularly true for blue skies research, which refers to curiosity-driven basic research where practical, real-world applications are not immediately evident, making it challenging and often unpredictable to assess its value (Florio and Sirtori, 2016).

The idea behind the MPC is simple. Unlike other markets, where the producers and consumers of goods are typically distinct, in the research market, the scientific community acts as both the producer and consumer of publications. Consequently, the implicit price the scientific community is willing to pay for an additional publication is expected to be at least equal to the production cost. A scientist's decision to dedicate time to conducting research, writing a paper, and ultimately publishing it requires them to allocate time to these activities, often at the expense of other professional commitments, such as teaching or administrative duties. As a result, the marginal production cost of a scientific publication in a year is determined by the author's annual shadow wage relative to her productivity (i.e., the number of publications in that year), adjusted by the proportion of time dedicated to research (European Commission, 2014).

---

10 See Morretta et al., 2022, Rosseau et al., 2021, Florio and Sirtori, 2016 for a discussion about the issue.

In empirical applications, the shadow wage is often proxied by the observed gross salary (Morretta et al., 2022; Bastianin et al., 2022; Bastianin and Florio, 2018; Florio et al., 2016). Practically, if a researcher has a gross annual wage of EUR 60,000 and uses 100% of her time on research, with a productivity of two scientific products per year, then the value of a publication would be estimated at EUR 30,000. Instead, if the researcher devotes only 50% of the time to the research, then the MPC would return a value of EUR 15,000 euro (Eq. 1) (see Appendix A for the full derivation of the MPC formula).

$$MPC_t = MPC_{res_t} = \left( \frac{w_{res_t} * h_{res_t}}{y_{res_t}} \right) \tag{1}$$

In Eq. 1, the average total marginal cost of producing a scientific publication ($MPC_t$) at time $t$ equals to the average marginal cost of doing research ($MPC_{res_t}$). $w_{res_t}$ denotes the annual salary of the author doing research; $h_{res_t}$ denotes the share of working time author dedicates to research; $y_{res_t}$ measures the author productivity in the period $t$, measured as the number of scientific products she produces in that period.[11] $t$ denotes time, generally the year. - It is omitted afterwards for simplicity.

Florio et al. (2016), echoed by Florio (2019), Bastianin and Florio, 2018; and Bastianin et al. (2022) applied Eq. 1 to value the scientific impact of the CERN LHC and CERN HL-LHC obtaining an MPC of EUR 11,011 per each scientific product (articles and pre-prints) generated by the LHC research. Using the same approach, Battistoni et al. (2016) obtained a value ranging from EUR 8,000 to EUR 8,250 (see **Table 1**). Eq. (1) has been subject to various criticisms (Rosseau et al., 2021), partially addressed by Morretta et al. (2022). First, the formula assumes full authorship, treating all scientific products as if they are authored by a single individual. This neglects co-authorship, leading to the problematic implication that a single-author publication incurs the same production cost as one written in collaboration with hundred or even thousands of co-authors. Second, additional relevant costs beyond wages, such as software for the analysis or costs for publishing services (e.g., editing, peer-reviewed process, etc.), are not accounted for and need to be included.

Third, the $MPC$ is an increasing function of wages, so publications from countries where wages are higher would be worth more, and therefore Purchasing Power Parity (PPP)-like correction is suggested to minimise the bias (Rosseau et al., 2021).[12] In addition, Eq. 1 is an inverse function of productivity ($y_{res}$) with the consequence that – if not appropriately estimated – publications from researchers who produce fewer publications over a given period would hold more value than those who generate more publications in the same timeframe.[13] Florio et al. (2016), Bastianin and Florio (2018), and Bastianin et al. (2022) arbitrarily set the productivity at 3.5 papers per author per year and keep it fixed over long time horizons (**Table 1**), disregarding the extremely high productivity in physics (Ioannidis et al., 2023), also as a consequence of research collaboration and its dynamics over time (see Section 5.1). Indeed, the literature argues that productivity is not fixed but is a function of the number of co-authors: it is reasonable to expect that the number of papers produced by a scientist in a given year is higher when working in collaboration (multi-authorship) than alone (Yadav et al., 2023; Abramo et al. 2017; Ductor, 2015; Gomez, 2015; Kato, and Ando, 2013). In addition to these adjustments, Eq.1 also requires correction to account for the quality and type of publication, such as considering the number of citations or similar quality metrics. Morretta et al. (2022), partially address the above criticisms by expanding Eq. 1 to incorporate the cost and business model in scientific research publishing (SQW, 2004) (see Appendix A for details). The authors focus on space research and propose the following formula (Eq. 2), where the average marginal cost of research ($MPC_{res}$) is augmented by the average marginal cost of publishing ($MPC_{pub}$):

$$MPC_t = MPC_{res_t} + MPC_{pub_t} = \left(\frac{w_{res_t} * h_{res_t}}{y_{res_t}}\right) + \left(\frac{w_{pub_t} * h_{pub_t}}{y_{pub_t}}\right) \qquad (2)$$

---

[11] When we have more authors, $w_{res_t}$ it their average wage, $h_{res_t}$ the average time they dedicate to research, and $y_{res_t}$ their average productivity. For instance, in the case of two authors, the first one with 4 publications in a year and the second one with 3, then their average productivity is 3.5. So, the average productivity is measured as by the ratio between the total number of scientific products produced by the authors in a year out of the total number of authors.

[12] A minor critic relates to the use of market wages as a proxy of shadow wages in regulated markets (as noted above).

Specifically, Morretta et al. (2022) improve Eq. 1 as follows:

- The annual (average) salary of the author(s) ($w_{res}$) in the first term of Eq. 2 is expressed in PPP in the case of authors from different countries taking on board the suggestion by Rousseau et al. (2021);
- The average researchers' productivity ($y_{res}$) attempts to address co-authorship by showing results either considering full or partial credit attribution. The full authorship attribution suffers from the same problems discussed for Eq. 1. Regarding partial credit attribution, the authors divide credits uniformly among the number of co-authors (e.g., Shen and Barabasi, 2014; Hirsch, 2007). In practice, if a paper has two co-authors, each one is assumed to contribute one-half and the parameter $y_{res}$ is set at 0.5. With the same logic, a paper with 100 co-authors (not a rarity in some scientific fields) results in a value of $y_{res}$ equals to 0.01. The implication is that the partial credit attribution always returns higher values of $MPC$ compared to the full credit attribution by definition, because the parameter $y_{res}$ is always smaller in the case of co-authorship. We further discuss the topic below.
- Regarding the second term, $w_{pub}$ is the average gross annual wage of editors and reviewers employed in the publishing process, $h_{pub}$ is the share of the time they employ in that process, and $y_{pub}$ is the yearly total number of papers they peer review, and publish. Overall, the $MPC_{pub}$ is valued at EUR PPS 484 in Morretta et al. (2022).[14]

---

[13] To be noted that Eq. (1) does not foresee any boundaries for the $MPC_t$, which can vary from zero to infinite according to the productivity value. Specifically: $\lim_{y_{res} \to 0} \left(\frac{w_{res}*h_{res}}{y_{res}}\right) = \infty$ and $\lim_{y_{res} \to \infty} \left(\frac{w_{res}*h_{res}}{y_{res}}\right) = 0$.

On average, depending on the method used to attribute publications to authors, the value of a $MPC_{res}$ ranges from EUR PPS 4,885 to 17,699. Higher values are associated with lower values of $y_{res}$ in multi-author papers captured by the partial credit attribution method. For publishers, the value of their activities is EUR PPS 484, resulting in a total $MPC$ ranging from EUR PPS 5,369 to 18,153. **Table 1** summarises the existing evidence. It is important to note that all the listed studies treat the scientists' productivity as a fixed parameter over long time periods.

**Table 1. Parameters and the estimate monetary value of scientific products in existing literature using the MPC approach.**

| Study | Field of application | Time horizon | Credit attribution method | $w_{res}$ EUR | $h_{res}$ (%) | $y_{res}$ Papers per author per year | $MPC_{res}$ EUR | $MPC_{pub}$ EUR | Total $MPC$ EUR |
|---|---|---|---|---|---|---|---|---|---|
| • Florio et al., (2016) <br> • Bastianin and Florio (2018) <br> • Bastianin et al. (2022)* | High-Energy-Physics – HEP (CERN LHC and HL-LHC) | Florio et al (2016): 1993 - 2025; Bastianin and Florio (2018) and Bastianin et al.(2022): 1993 - 2038 | Full | 52,289[a] | 65% | 3.5 | 11,011[a] | n/a | 11,011[a] |
| Battistoni et al. (2016) | Medical research (National Hadrontherapy Centre for Cancer Treatment in Italy – CNAO) | 2002 - 2031 | Full | 50,000 – 55,000[b] | 30% - 80%[c] | 2 - 5[d] | 8,000 – 8,250[e] | n/a | 8,000 – 8,250[e] |
| Morretta et al. (2022) | Space research (Italian Space Agency) | 1998 - 2018 | Full | 40,204[f] | 50% | 4.0[f] | 4,885[f] | 484.25[f,g] | 5,369[f] |
| | | | Partial | 40,024[f] | 50% | 1.1[f] | 17,669[f] | 484.25[f,g] | 18,153[f] |

*Source Authors. Note: parameters' subscripts omitted for simplicity. *Bastianin and Florio (2018) and Bastianin et al. (2022) projected the analysis of Florio et al., (2016) up to 2038, without changing the fundamental parameters of the model, including those*

---

[14] It refers to the average marginal cost of publishing for peer-reviewed papers in journals and books (Morretta et al. (2022: Table 9). The authors also provide the average cost of publishing of conference proceedings set at EUR PPS 158.94. For the sake of simplicity, we report in **Table 1** only the first value.

*in the table above.* $^a$ *EUR 2013;* $^b$ *55,000 for CNAO insider scientists and 50,000 for extra-CNAO scientists. EUR 2013* $^c$*A triangular distribution is assumed with a modal value of 30% for CNAO insider scientists and 80% for extra-CNAO scientists (Battistoni et al., 2016:86).* $^d$ *A normal distribution with a standard deviation of 0.3 around the mean value of 2 papers per year for CNAO insider scientists and of 5 papers per year for extra-CNAO scientists (Battistoni et al., 2016:86);* $^e$*EUR 2013;* $^f$*Yearly average values (1998-2018).* $^g$ *It is the value related to the $MPC_{pub}$ of scientific publications in journals and books. Morretta et al. (2022) also report the $MPC_{pub_t}$ of conference proceedings amounting to EUR PPS 158.94.*

## 3. Method

This paper focuses on $MPC_t$ in Eq. 2 and specifically on the authors' productivity ($y_{res_t}$) in case of co-authorship, which is used as a proxy for collaboration (Yadav et al., 2023; Abramo et al., 2011). We treat the marginal cost of publishing ($MPC_{pub_t}$) as an exogenous parameter, as it is not the focus of our research and is not expected to vary significantly based on whether the publication has multiple authors.

We argue that existing applications fail short on two key points. First, they treat the authors' productivity ($y_{res_t}$) as a fixed parameter; however, we argue that it is an increasing function of collaboration activity. There is longstanding evidence showing a strong link between research collaboration and publishing productivity (Yadav et al., 2023; Abramo et al. 2017; Ductor, 2015; Gomez, 2015; Kato, and Ando, 2013; Abbasi et al., 2011; Abramo et al. 2011; 2009; McFadyen and Cannella, 2004). For instance, Abramo et al. (2009) focus on the Italian academic research system at the institutional level and find that, on average, extramural collaboration intensity is significantly correlated with productivity, particularly in fields such as industrial and information engineering, biological sciences, and agricultural and veterinary sciences. In physics, productivity is strongly correlated with international collaboration. The positive relationship between research productivity and the degree of international collaboration achieved by a scientist is also documented in Abramo et al., (2011). Kato and Ando (2013) replicate these findings in chemistry, while Ebadi and Schiffauerova (2016) demonstrated that larger research teams tend to produce more publications. Additionally, Lee and Bozeman (2005), using curriculum vitae and survey data from 443 US-based academic scientists, show a significant correlation between the number of collaborators and the volume of peer-reviewed papers. More recently, Yadav et al. (2023) analysed data from three small economies (Ireland, Denmark, and New Zealand) and demonstrated that collaborating with highly-cited 'star' scientists (a scientist who is at or above the 95th percentile of scientists in the cumulative distribution of citations received) significantly boosts productivity.

Second, while we agree with Moretta et al. (2022) that the partial credit attribution is a valid approach for valuing scientific publications, the measurement and treatment of authors' productivity need to be handled more carefully. We argue that Morretta et al. (2022) misinterpret the role of the productivity parameter $y_{res_t}$ in their analysis. In the partial credit attribution method, that parameter reflects the individual contribution of each co-author to a publication rather than their actual productivity. Indeed, the authors' productivity of 4 scientific products per year, as calculated in their paper, is a stylised fact (see Table 6 in Morretta et al., 2022) and should not change based on the attribution method used. The misinterpretation leads to puzzling implications, such as the claim that "productivity" decreases as collaboration or multi-authorship increases, which contradicts the available evidence. As Morretta et al. (2022) state "*By crediting each co-author with (only) a share of publication, the total number of publications released by all the authors is lower than in the full credit attribution method. This leads to lower productivity that, subsequently, increases the marginal cost value.*" However, the total number of publications is a fact and should not vary change depending on the estimation method. Instead, the models should be designed to explain the observed data.

To solve these issues, we propose Eq. 3, which integrates Eq. 2 with a simple, but effective adjustment to value multi-author scientific publications:[15]

$$MPC_{f,t} = MPC^c_{res_{f,t}} + MPC_{pub_t} = \left(\frac{w_{res_{j,f,t}} * h_{res_{j,f,t}}}{y_{res_{j,f,t}}(n_t)}\right) * n_t + MPC_{pub_t} \qquad (3)$$

Where now the $MPC_{f,t}$ is the average marginal cost of producing a generic scientific publication in the scientific field $f$ at time $t$ and $MPC^c_{res_{f,t}}$ is the average marginal cost of doing research in that field. $w_{res_{j,f,t}}$ denotes the average annual salary of the $j-th$ author(s) doing research $with\ j = 1,..,n,...,N$; $h_{res_{j,f,t}}$ denotes their average share of working time dedicated to research and it ranges from 0 to 100%. In this model, the author(s)' productivity is now a function of the (average) number of co-authors ($n_t$), which is also used as a scaling factor for the term $MPC^c_{res_{f,t}}$ to incorporate co-authorship into the equation. The superscript "c" stands for co-authorship (collaboration) and differentiates this new term from $MPC_{res_t}$ in Eq. 2. For simplicity, $MPC_{pub_t}$ is not explicitly included in Eq. 3, but a brief discussion of this parameter is provided in Appendix B. The interpretation of $MPC^c_{res_{f,t}}$ is straightforward: given the time allocated to research ($h_{res_{j,f,t}}$), co-authorship reduces each author's individual contribution to a paper (as in Morretta et al., 2022), but it increases productivity, allowing for more articles to be produced in a given period. The term in brackets $\left(\frac{w_{res_{j,f,t}} * h_{res_{j,f,t}}}{y_{res_{j,f,t}}(n_t)}\right)$ represents single-author publications, and therefore requires adjustment for the number of co-authors. The functional form of $y_{res_{j,f,t}}(n_t)$ is unknown a priori and may vary by scientific fields, time period, and other factors, as collaboration practices differ across disciplines (Abramo et al., 2017; Abramo et al., 2009) and evolve over time (Abrahams et al. 2019; Abbasi et al., 2011).

---

[15] One may argue that scientific collaboration generates other costs than salaries and publishing. International collaborative activities also introduce logistical (travel costs) and coordination challenges that can increase the time (and production costs) needed to complete a publication or project (Jones et al.2008; Wagner, and Leydesdorff, 2005). Assuming that these costs can be calculated, they are not publications-specific, and the challenge is to factorise and attribute them to the specific scientific product. Similarly to the cost of publishing, these costs only have a limited role (if any) compared to the cost of research.

Often, authors working in a given scientific field produce different types of scientific products (conference proceedings, peer-review articles, notes, pre-prints and so on) that may cover different topics.[16] When Eq.3 is applied to the entire catalogue of scientific products, it returns the unitary average marginal production cost of publications in the scientific field $f$ as a function of the number of coauthors without distinguishing by the type of scientific product. Instead, one may be interested in the monetary value of a publication covering a given topic, a specific type of publication (e.g. per-review articles) or both. This necessity may arise especially when one needs to attribute in a causal sense the monetary value of scientific products to a specific project or experiment, which is part of a wider research programme. For instance, in our case, the value of the scientific production attributable to the LHC scientific programme as part of the CERN research activity. In that case, one might be tempted to adjust Eq.3 by integrating a reduction factor ($\alpha$) to factorise the share of research time author(s) dedicate to research ($h_{res_{j,f,t}}$) to that particular scientific product. To give an example, suppose an author in the HEP scientific field produces three articles in a year: two dealing with quantum computing applications and one covering particle collision simulations at the LHC, with all the other parameters being equal. One may want to evaluate the article on LHC event simulations by reducing the total value of the author's research to one-third by introducing a reduction factor ($\alpha_i = 0.33$) to account for the proportion of research time she dedicated to that product over the entire her catalogue. We demonstrate that this approach introduces puzzling implications in the analysis and contradictory evidence (see Appendix C). Instead, we argue that the number of co-authors is still a valid (although imperfect) criterium to discriminate between sub-set of scientific products and that the productivity function already incorporates the necessary information to do it. We use the CERN research and the LHC scientific programme, along with its related bibliographic records, as a case study to calibrate the parameters entering Eq. 3, show how to value publications arising from collaboration in HEP, and discuss the results, including limitations.

## 4. Collaborative scientific publication at the LHC

Scientific collaboration at CERN is highly organised, international, and characterised by a multidisciplinary approach that allows for the execution of some of the most complex scientific experiments in the world. CERN's primary research focus is on particle physics, carried out through massive experiments like ATLAS, CMS, ALICE, and LHCb, all based at the Large Hadron Collider (LHC). These experiments involve collaborations among thousands of scientists, engineers, and technicians from over 100 countries, and institutions worldwide. Collaborations are organised into international teams, each responsible for different aspects of the experiments, such as detector construction, data analysis, and software development.

---

[16] For instance, in our case we will distinguish LHC-related research topic from the other topics studied by our sample of authors in HEP. Other examples can be the following ones. Under the climate science, research topics can cover climate system dynamics and modelling, cryosphere studies or greenhouse gas emissions and carbon cycling. In health, research topics can deal with epidemiology and infectious diseases, aging and geriatrics, chronic diseases, and many others.

CERN promotes open data and shared resources by providing access to experimental data through platforms like the CERN Open Data Portal. Resources such as computing power are shared via the Worldwide LHC Computing Grid (WLCG), a distributed infrastructure involving over 170 computing centres in 42 countries. Large-scale infrastructure, such as accelerators, detectors, and computing facilities, are shared among the participating institutions. This pooling of resources allows for experiments on a scale that no single country could afford alone.[17]

In this role, CERN serves as the central hub, coordinating activities, managing infrastructure, and facilitating smooth collaboration among diverse international teams. Researchers from these international teams regularly meet to discuss progress, share results, and plan future steps. This includes daily working group meetings and larger collaboration meetings held several times a year to ensure alignment and continuity in their scientific efforts. Scientific publications at CERN are the result of a highly collaborative process. Typically, research begins with small groups or subgroups of a collaboration, where specific data are analysed or new ideas tested, laying the groundwork for a potential publication. Before any result is considered for publication, it undergoes an internal review process within the collaboration, including internal presentation, review committees, and cross-checks. Once the analysis is validated, a small group of scientists typically drafts the manuscript. This draft includes detailed descriptions of the methodology, data analysis, results, and interpretations. After the draft paper is prepared, it is circulated among all collaboration members for comments and feedback, a process that can involve from hundreds to thousands of co-authors, given the large scale of the collaboration. This iterative review process ensures that the paper reflects the collective expertise and consensus of the entire collaboration (Ellis, 2020; Voss, 2012).

Authorship in CERN papers can be extensive. For major experiments, the list of authors can include several thousand names, representing all or a subsample of scientists who have contributed to the experiment during a certain period. The names are usually listed alphabetically by surname, without distinguishing between primary authors and contributors. This egalitarian approach recognises the collaborative nature of the work. On top of that, each author's institutional affiliation is listed, demonstrating the global and multidisciplinary nature of the collaboration (Pritychenko, 2016; Castelvecchi, 2015).

Once the collaboration approves the final manuscript, also via specific committees, it is submitted to a peer-reviewed scientific journal. Many of its publications are made freely available through open-access repositories such as arXiv or the CERN Document Server (CDS) to ensure that CERN-related publications are accessible to the global scientific community.[18] The CERN collaboration model aims to foster innovation, accelerate scientific discovery, and push the boundaries of human knowledge.

---

[17] See https://home.cern/about/who-we-are/our-governance for more information on the financial structure of CERN and its budget.
[18] See also the SCOAP3 as mentioned above.

# 5. Data
## 5.1 Search strategy and descriptive statistics

We extracted data from the INSPIRE HEP database covering the period 1990-2021.[19] INSPIRE HEP is the leading platform for scientific content in HEP as it aggregates content from multiple authoritative sources in this field, including ArXiv, CERN, DESY and a wide range of scientific publishers (e.g., Elsevier, Springer).[20] The data extraction process yielded 54,384 LHC-related identifiable individual authors who authored 434,065 unique scientific products (see below), of which 8% (n = 34,558) were LHC scientific products (hereafter LHC-related scientific products) and the remaining 92% (n = 399,507) non-LHC products, including other CERN collaborations/experiments scientific outputs and non-CERN related ones (hereafter non LHC-related scientific products). LHC-related scientific products were from the main LHC experiments, such as ALICE, ATLAS, CMS and LHCb, but we also considered smaller experiments and collaborations as part of the scope of our research (Table 2).[21] The data collection process is outlined below, with further details provided in Appendix D to facilitate replication of our analysis in other scientific fields beyond HEP. All scripts for data collection, processing, analysis, and visualisation were developed in R and are available as Supplementary Materials for this paper.

**Table 2. Experiments and collaborations related to the LHC considered in this paper**

| Collaborations | Experiments |
|---|---|
| - ALICE<br>- ATLAS<br>- CMS<br>- FASER<br>- LARP<br>- LHCb<br>- LHCf<br>- MATHUSLA<br>- milliQan<br>- MoEDAL<br>- RD12<br>- RD2<br>- RD23<br>- RD42<br>- RD5<br>- RD8<br>- SND@LHC<br>- TOTEM<br>- USCMS | - AcerMC<br>- CERN-H4IRRAD<br>- CERN-HL-LHC<br>- CERN-LHC<br>- CERN-LHC-ALICE<br>- CERN-LHC-ATLAS<br>- CERN-LHC-CMS<br>- CERN-LHC-FASER<br>- CERN-LHC-FELIX<br>- CERN-LHC-FP420<br>- CERN-LHC-HV-QF<br>- CERN-LHC-LHCb<br>- CERN-LHC-LHCf<br>- CERN-LHC-MOEDAL<br>- CERN-LHC-TOTEM<br>- CERN-RD-002<br>- CERN-RD-005<br>- CERN-RD-008<br>- CERN-RD-012<br>- CERN-RD-013<br>- CERN-RD-018<br>- CERN-RD-023 |

---

[19] The timeframe for this analysis is justified by key milestones in the development of the LHC. The LHC's construction phase was formally approved in 1993, and operations commenced in 2008. However, the early 1990s marked the beginning of significant scientific production related to its development, which requires the inclusion of this period in the analysis. The endpoint of 2021 was selected based on the availability of comprehensive data through the INSPIRE API up to that year at the time of data extraction (see Appendix A.2). Six LHC-related papers were identified in the period 1983–1989 period, they were excluded from the analysis.
[20] See https://help.inspirehep.net/knowledge-base/inspire-content-sources/ . Last access in October, 2024.
[21] This represents an additional step forward compared to Florio et al., (2016), Bastianin and Florio (2018), and Bastianin et al., (2022), which have a narrower scope The analysis by Florio et al (2016) relies on a total of about 23,000 scientific products, including projections considering the period 1993 – 2025 obtained looking for only on a subset of experiments/collaborations (see Carrazza et al., 2014 for details).

| | - CERN-RD-027
- CERN-RD-042
- CERN-RD-049
- FNAL-E-0892
- FNAL-E-0893
- LARP
- MATHUSLA
- milliQan
- SND@LHC |
|---|---|

*Source:* Authors based on INSPIRE HEP database, July and August 2024.

For each author, we collected their entire catalogue of scientific documents, spanning from 1990 to 2021. This catalogue includes both LHC and non-LHC-related scientific products, such as outputs from other CERN experiments/collaborations and non-CERN publications. All types of scientific products were considered, including peer-reviewed research articles, pre-prints, conference proceedings and papers, notes, theses, books and books chapters, and reports. HEP pre-prints often become publications later in the research process, either in proceedings (books or journals) or peer-reviewed research journals. Also, since a very long time, experimental HEP proceedings material is only partially original material. A good part, especially when only presented by one author ("on behalf of the X collaboration"), of this material will be found again in a research paper at some stage. Accordingly, we removed duplicates in our database using the unique control number variable.[22] Moreover, INSPIRE HEP database updates the record once a pre-print gets published.

Then, for each collected scientific product, whether LHC or non-LHC-related, we recorded the following key information: title, earliest year of record in the INSPIRE HEP database, number and list of co-authors, type of products (e.g., articles, pre-prints, books, books chapters, conference proceeding, and so on), number of citations, and the unique identifier of the product. We also identify authors based on a unique code, including the full name and the RECID, i.e., the primary unique identifier of each record in the INSPIRE HEP database, allowing unambiguous attribution of publications to authors. Finally, for each author we collected information on their affiliation and country, based on institution affiliation (Yadav et al., 2023).

**Figure 1** shows the scientific outputs of the authors in our sample, distinguishing between LHC-related (blue line) and non-LHC-related products (red line). Overall, it documents a steady increase in scientific production over time, moving from 7,355 scientific products in 1990 to an average of 17,168 products per year in the period 2008 – 2021. As expected, the trend for LHC scientific products aligns closely with the LHC's lifecycle and major milestones. LHC's scientific production increased steadily during the construction phase (1993 – 2007) and peaked during the operating phase. Two notable spikes occur: the first in 2012, coinciding with the discovery of the Higgs boson (2 366 products), and another in 2016 (2 343 products), marking the beginning of the second operating phase after a long shutdown for upgrades (2013–2015). These peaks reflect the major advancements and breakthroughs achieved by the LHC.

---

[22] This process may still result in some negligible number of duplicates, though they are not readily identifiable.

Parallel to LHC-related production, the scientists involved in LHC experiments also contributed to non-LHC-related scientific output. If considering the total time period, non-LHC scientific production (red line) is more than nine times higher than LHC-related output (as shown on the right-hand side Y-axis), which can be explained by the fact that many scientists only spend part of their careers working on LHC experiments, and the sample includes authors who were not continuously involved with the LHC. [23]

**Figure 1. Number of scientific products by scientists in our sample. Total production, LHC- and non-LHC-related scientific production (1990-2021).**

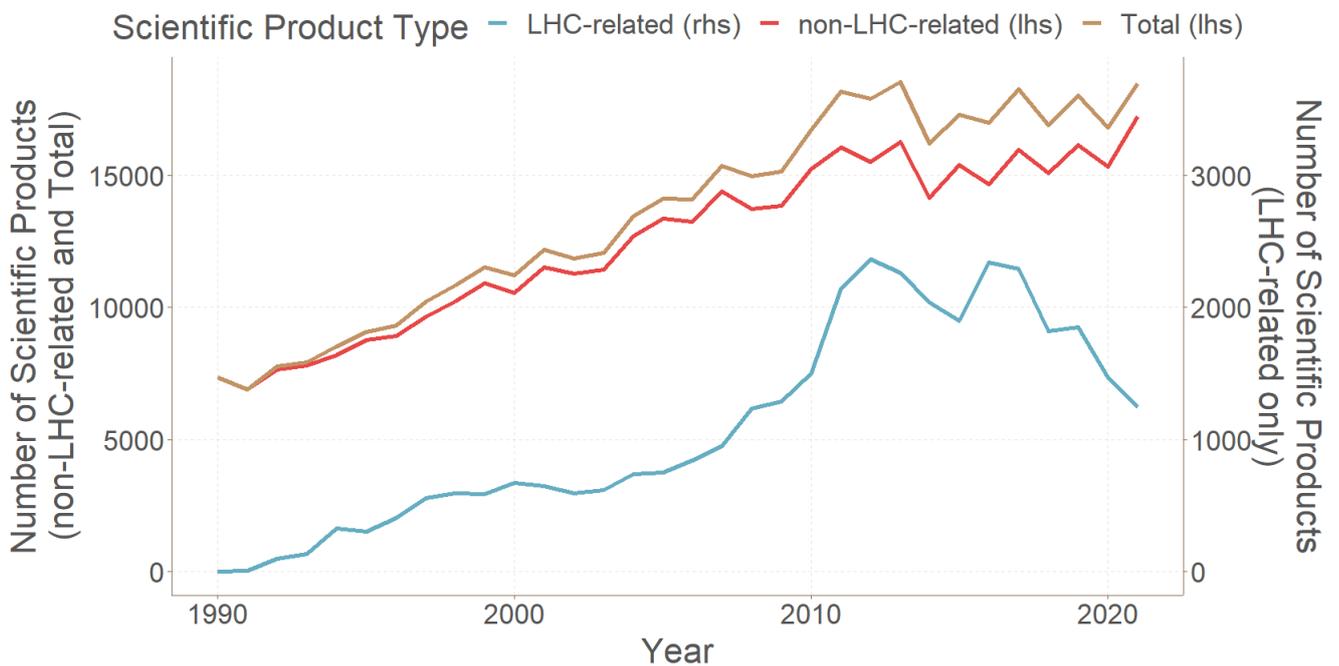

*Source*: Authors based on INSPIRE HEP data. Full sample of 434,065 scientific products produced by authors with at least one LHC-related scientific product in 1990-2021. *Note:* Scientific Product type based on LHC experiments and collaborations. Years corresponding to earliest date recorded in INSPIRE HEP database.

---

[23] We identified them as authors with zero LHC publications for several years.

The sample of scientific products produced by the authors is extensive and includes a diverse range of outputs (**Figure 2**), Articles and pre-prints represent more than half of the sample (54.3%) when considering the entire sample of scientific products, while they are 16.3% of all the LHC-related products. As mentioned above, it is common practice among authors in high-energy physics (HEP) to use proceedings to present original material on behalf of LHC Collaborations or Experiments, which often results in research papers at some stage. This approach explains the high share of conference proceedings and papers (56.7%) over the total number of LHC-related scientific products. In contrast, books, book chapters, theses, notes, and reports account for a smaller share, either considering the total number of scientific outputs or when restricting the sample to the LHC-related products.

**Figure 2. Scientific products by type (1990 – 2021)**

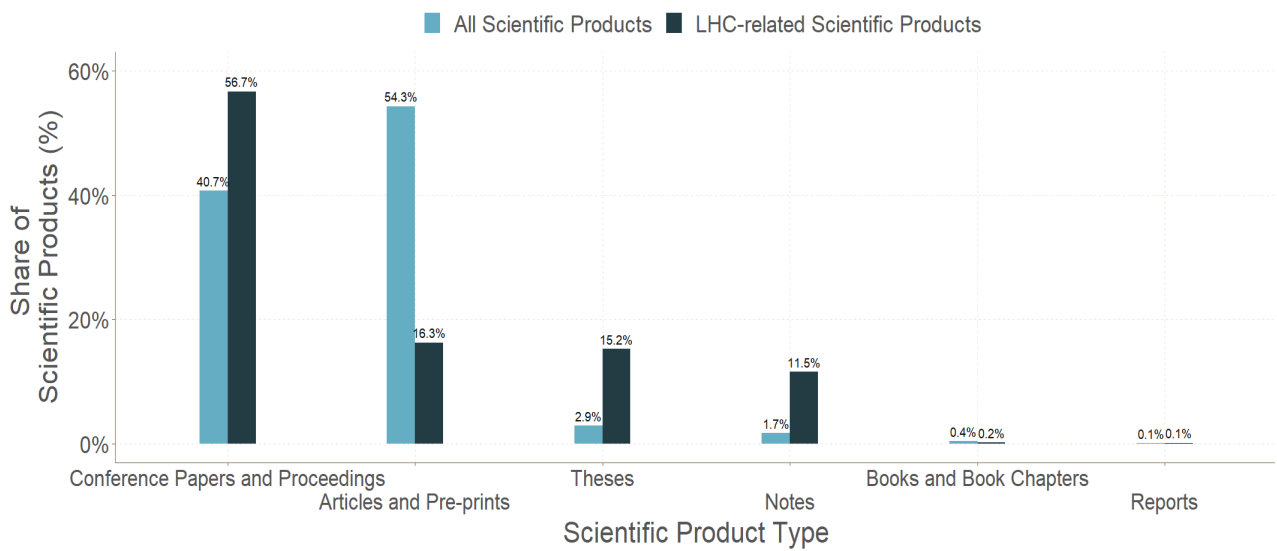

Source: Authors based on INSPIRE HEP data. Sample of 431 935 scientific products in 1990-2021 where the type of document was unambiguously retrievable. Products include both LHC-related and on-LHC-related ones. Note: We grouped books and book chapters, conference papers and proceedings, which were individual types in the original data.

**Figure 3** shows the distribution of all the scientific products by number of (co-) authors over time. More than 75% of the products are co-authored by at least 2 authors, and specifically by 3-5 co-authors (30%), 6-10 (about 10%), and 11-25 co-authors (about 8%). The share of products with a higher number of authors has grown over time following the global trend documented in Section 1, but remained quite limited. Large co-authorships with more than 100 authors mainly characterise report writing (**Figure 4**). Statistics change dramatically when we look at the LHC-related products (**Figure 5**). Interestingly, scientific products with an extremely high number of coauthors (more than 100) have surged since 2008, when the LHC began its research activity (green bars in **Figure 5**) and account for 20-25% of all the LHC-scientific production in more recent years. These products are primarily research articles, pre-prints and reports (**Figure 6**). Over 75% of articles and pre-prints had more than 100 co-authors between 2008 and 2021. The percentage is even higher for reports.

**Figure 3**. Percentage of total scientific products by number of co-authors (1990 – 2021)

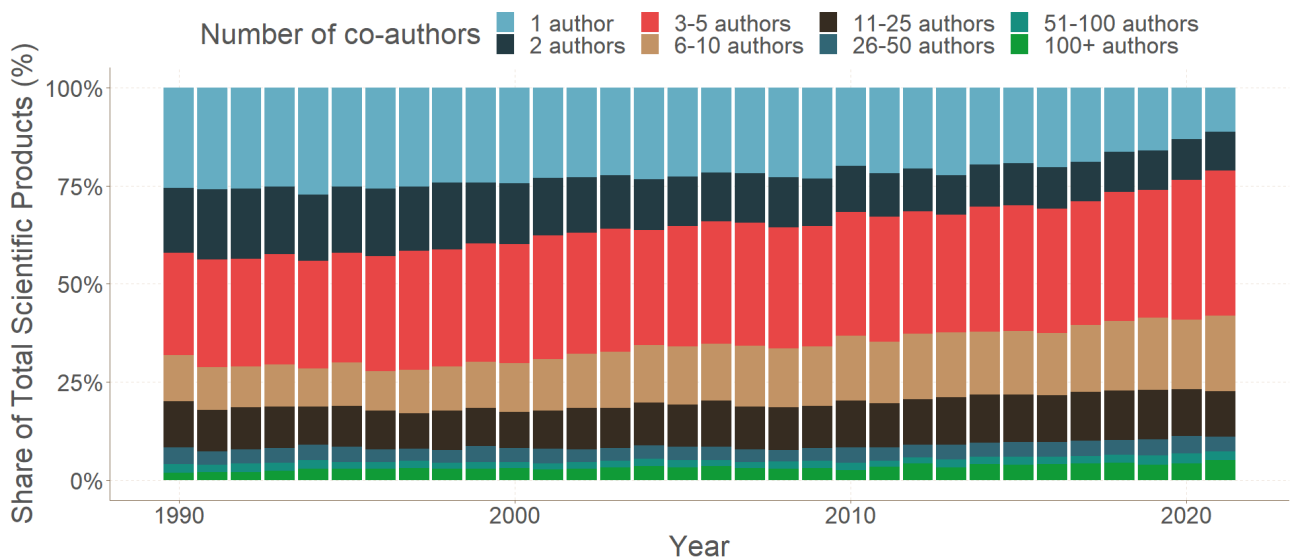

Source: Authors based on INSPIRE HEP data. Full sample of 434,065 scientific products involving the identifiable authors with at least one LHC-related scientific product during the 1990-2021 period.

**Figure 4**. Percentage of scientific products by type and number of co-authors (1990 – 2021)

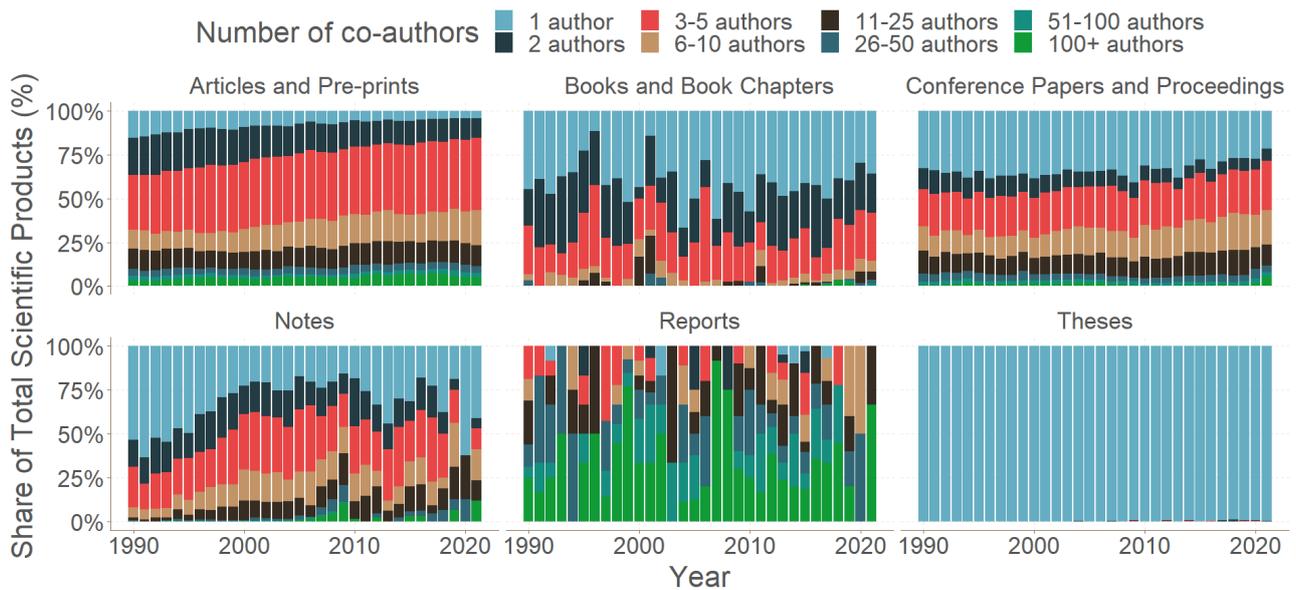

Source: Authors based on INSPIRE HEP data. Full sample of 431 935 scientific products with a unambiguous document type during the 1990-2021 period. Note: Classes of numbers of authors determined based on statistical analysis of their distribution and rounding.

**Figure 5**. Percentage of LHC-related scientific products by number of co-authors (1990 – 2021)

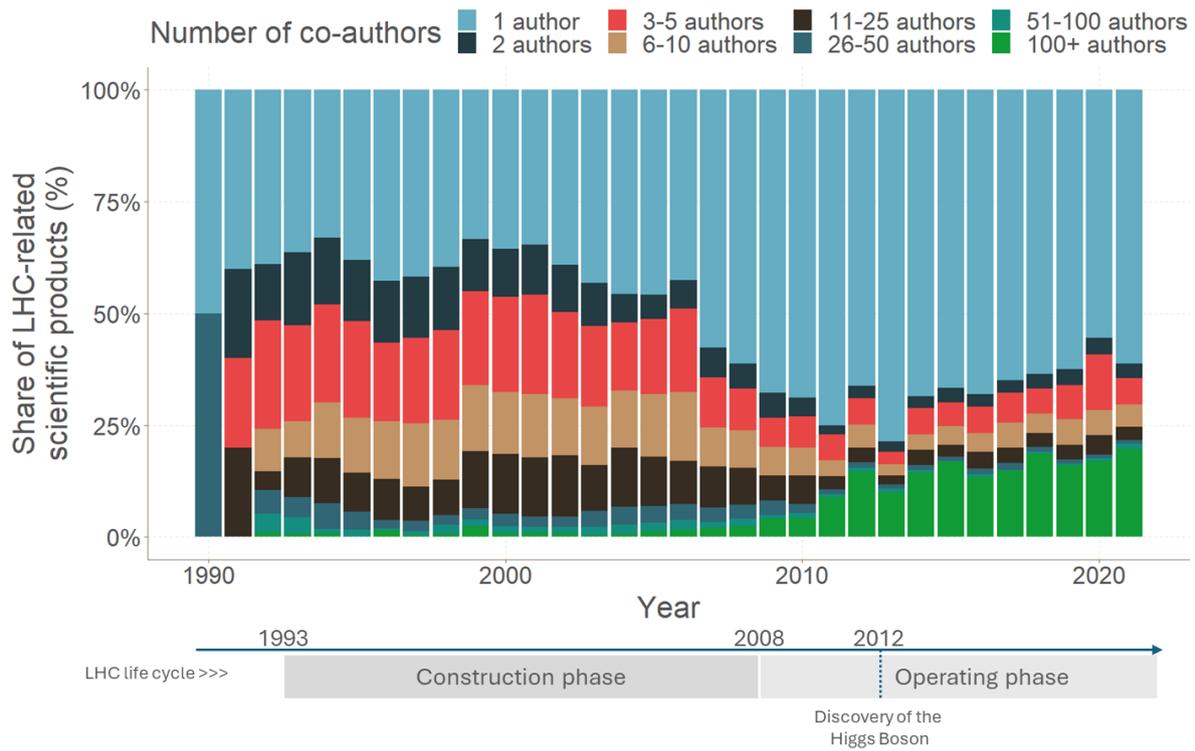

Source: Authors based on INSPIRE HEP data. Full sample of 34,558 LHC-related scientific products with identifiable authors during the 1990-2021 period.

**Figure 6**. Percentage of LHC-related scientific products by type and number of co-authors (1990 – 2021)

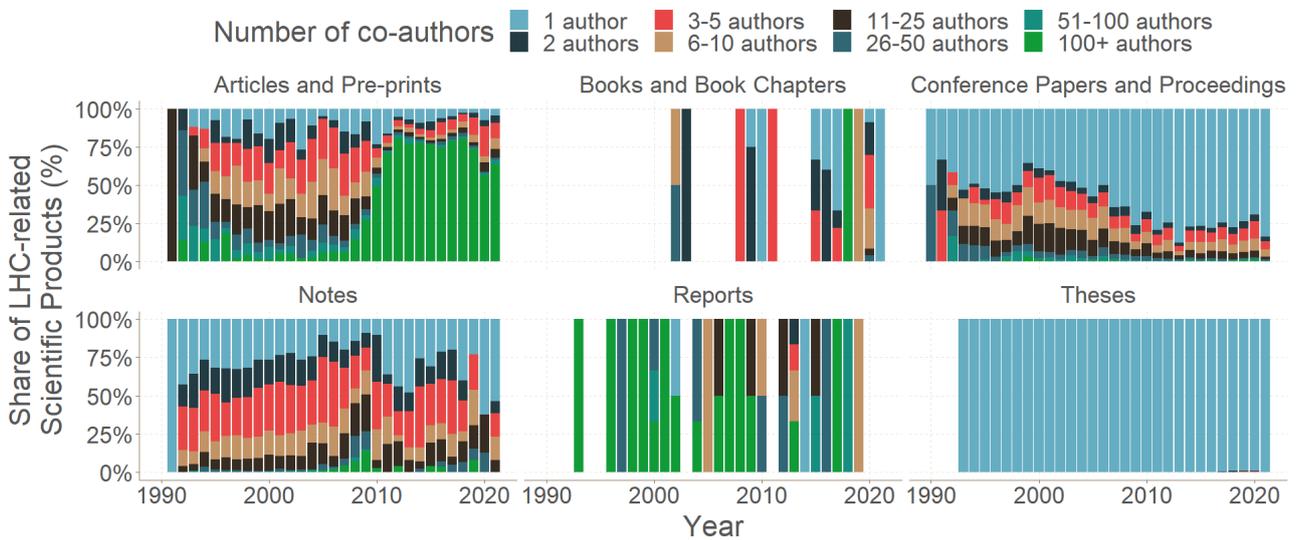

Source: Authors based on INSPIRE HEP data. Full sample of 33,573 LHC-related scientific products with a unambiguous document type during the 1990-2021 period. Note: Classes of numbers of authors determined based on statistical analysis of their distribution and rounding.

## 5.2 Scientists' productivity

The calculation of the marginal cost of scientific products needs to assess the yearly authors' productivity, defined as the number of scientific products per year produced by a scientist involved in the LHC research programme in her career, including her total scientific production, i.e. related and non-related to the LHC. Differently from previous papers and in accordance with the literature, we argue that collaboration strongly influences scientists' productivity (Section 3). To avoid bias caused by outliers, we set an upper limit of 3,000 co-authors for our sample. Consequently, 162 scientific products with more than 3,000 authors were removed, which accounted for only 0.04% of the total sample of products. All the subsequent analyses in this paper are based on this trimmed sample.[24]

**Figure 7** shows a significant increase in the yearly scientific output of scientists since 2008. The average output per author started at 5.9 products per year in 2008, rising to a range of 18 to 32 products per year between 2012 and 2021.[25] The increase in productivity is mainly attributable to an increase in LHC-related scientific production, as illustrated in **Figure 8**.

**Figure 7. Total productivity of authors (1990 – 2021).**

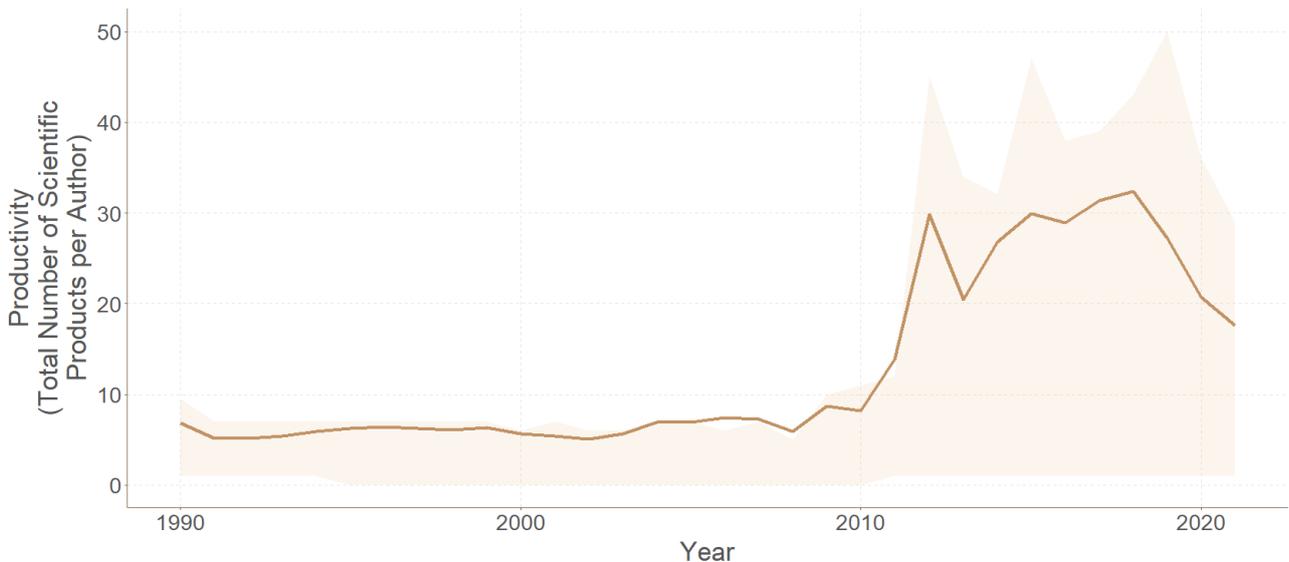

Source: Authors based on INSPIRE HEP data. Trimmed sample of 52,933 authors with at least one LHC-related scientific product in their catalogue. Note: The bold line corresponds to the mean productivity by author, while the upper and lower bounds of the areas are the Q75 and Q25, respectively. Productivity considers both LHC-related and non-LHC-related scientific products. Products with more than 3,000 co-authors excluded Authors with no scientific products (LHC or non-LHC-related) that are considered active are counted in this metric.

---

[24] The trimmed sample consists of 52,933 authors, who authored 433,903 scientific products.
[25] The small decline in the later years is documented for all types of products (LHC-related or not), and could partially be linked to the time lag to register publications in the database.

**Figure 8. LHC-related scientific production (1990 – 2021).**

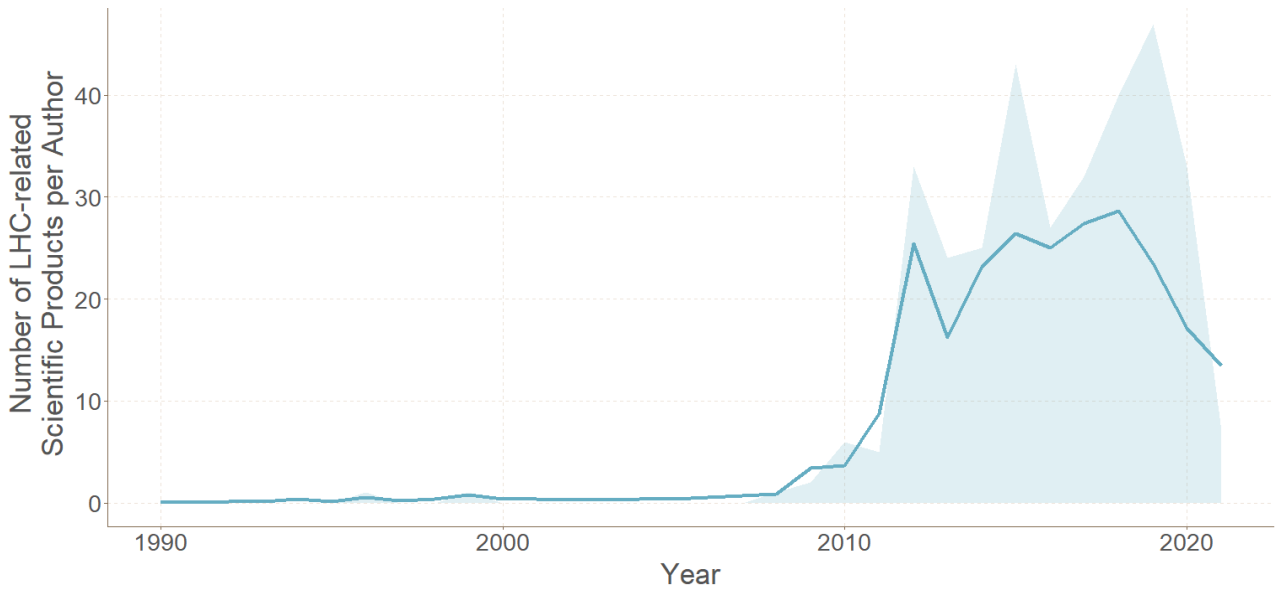

Source: Authors based on INSPIRE HEP data. Trimmed sample of 52,933 authors with at least one LHC-related scientific product in their catalogue. Note: The bold line corresponds to the mean number of LHC-related scientific products by author, while the upper and lower bounds of the areas are the Q75 and Q25, respectively. Products with more than 3 000 co-authors excluded. Authors with zero LHC-related scientific products in a given year are included in this metric.

The relationship between productivity and co-authorship, i.e. the function $y_{res_{j,f,t}}(n_t)$ in Eq. 3, is illustrated in **Figure 9.** There is a positive and statistically significant relationship between individual productivity and the number of co-authors. The Pearson correlation coefficient is 0.67, statistically significant at 1% level ($p\text{-value} < 2.2e\text{-}16$). The empirical relationship between productivity and co-authorship is not linear, though. We obtained an approximation of the function $y_{res_{j,f,t}}(n_t)$ by applying a non-parametric local regression model (Cleveland, 2017; Loader, 2006) visualised by the dark curve in **Figure 9.**[26] In our research, the functional from of $y_{res_{j,f,t}}(n_t)$ is linked to how the scientific activity of the HEP community at CERN is organised. However, different functional forms may emerge depending on the research field and the specific time horizon under analysis. Ignoring this stylised fact can lead to distorted productivity estimates and inaccurate assessments of the publication value as visualised in Table 1 in Section 2.

---

[26] The non-parametric regression method LOESS (Locally Estimated Scatterplot Smoothing) was used to estimate the productivity curve and capture the underlying trend in the data allowing for a flexible approximation of complex, non-linear relationships among data. Concretely, the curve was fitted using the sample of author-year productivity/mean number of authors pairs. The regression was implemented using the loess function of the stats R package, with the standard parameters (i.e., alpha parameter for smoothing of 0.75).

**Figure 9. Empirical relationship between individual productivity and co-authorship (1990 – 2021)**

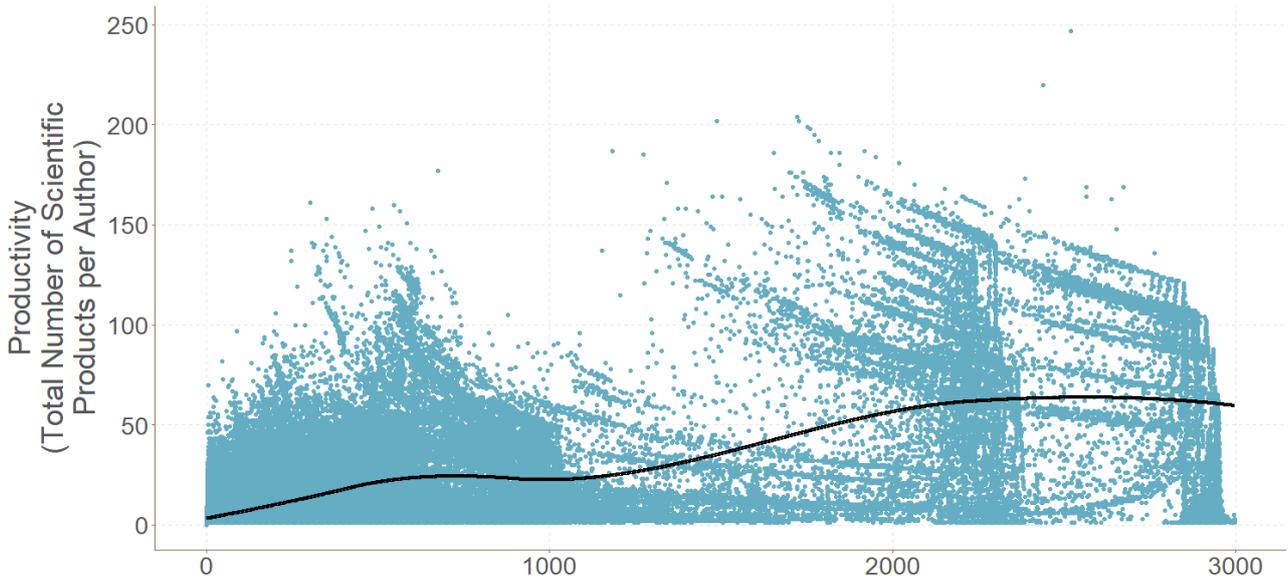

Source: Authors based on INSPIRE HEP data. Trimmed sample of 52,933 authors with at least one LHC-related scientific product during 1990-2021. Each point represents an individual author for a given year. Data contains both LHC-related and non-LHC-related scientific products (trimmed to exclude products with more than 3,000 co-authors). Data includes years where active researchers had zero scientific products. Note: curve fitted by LOESS regression, span parameter (alpha) of 0.75. This value is the standard for LOESS regressions and strikes a good balance between the use of sufficient data (given the high number of observations) and the risk of overfitting, as shown by the simple resulting curve (see Buchan, 2024).

### 5.3 Scientists' salary

The salary earned by scientists in the HEP community for writing their papers ($w_{res_{j,f,t}}$) in Eq. 3 is used to proxy the cost of producing a publication (Morretta et al., 2022; Florio and Sirtori, 2016). A scientist's salary may vary based on a set of individual traits (e.g., seniority) and contextual factors like affiliation and country. Our database of authors includes affiliation for 56% of authors (n = 21,500). If an author had multiple affiliations, we assigned them equally to her listed affiliation.[27] Using unique identifiers and affiliation data, we assigned each scientist to a country. Authors in our sample were based in institutions across 97 different countries, with the largest shares located in the USA (21%), Germany (11%), France (10%), Italy (9%), the UK (7%), and Switzerland (6%). Overall, 95% of authors were from institutions located in one of the 28 countries listed in **Figure 10** (from the USA to Austria). The remaining 5% were based in other countries across all continents, including European countries (e.g. Estonia, Romania, Bulgaria), South American (Paraguay, Brazil, Colombia, Trinidad and Tobago), Asian (Mongolia, Azerbaijan, Pakistan), African (Egypt, Togo, Ivory Coast) and many others.

---

[27] Attributing the author to her first listed affiliation do not chance the results in our analysis.

**Figure 10. Share of authors by country (n = 21,500) (1990 – 2021)**

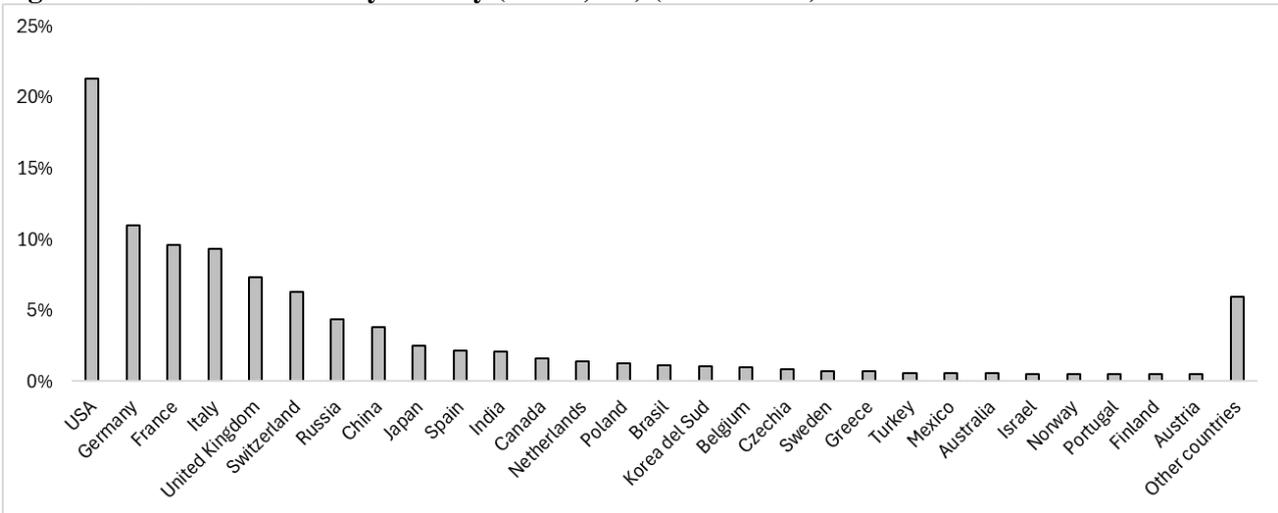

Source: Authors based on INSPIRE HEP data.

Since salaries depend on various factors, obtaining reliable global data on researchers' salaries presents significant challenges. International institutions such as the International Labour Organization (ILO) offer some data at the occupational and sectoral levels, but it lacks the granularity needed to accurately capture the heterogeneity of salaries across the global research community and, even more so, across time. Therefore, we rely on salary data provided by Morretta et al. (2022: Table 7) at the country level in EUR and in EUR PPS. The data fits our purpose for a number of reasons: (i) the data refer to the research sector, mostly academia; (ii) they cover almost all the countries of the LHC scientists in our database listed in **Figure 10** and cover 95% of our database; (iii) data are provided in time series covering the period from 1998-2018. We followed the methodology employed in Morretta et al. (2022) to expand the period back to 1990 and forward to 2021, and attribute salary to authors in the countries not covered by Morretta et al. (2022) but present in our list (Brazil, Canada, South Korea, Mexico, Russia).[28]

**Table 3** reports the scientists' average gross salary by country in the 1990-2021 period, while Appendix E reports the average gross salary by year and country in 2021, which is the most recent year in our database.

---

[28] We adopted country-specific Gross Domestic Product deflators provided by the World Bank, set 2006 as the base year, and computed salary values per each country for the entire period covered by our analysis 1990-2021.

**Table 3. Average gross salary in the research sector by country in the period 1990- 2021 (baseline year: 2006)**

| Country | Avg. gross salary | |
|---|---|---|
| | EUR | PPS |
| Australia | 64,607 | 61,202 |
| Austria | 63,062 | 60,145 |
| Belgium | 58,114 | 54,769 |
| Brazil | 8,079 | 25,730 |
| Canada | 84,609 | 81, 256 |
| China | 3,397 | 14,358 |
| Czechia | 19,356 | 35,776 |
| Finland | 46,071 | 37,229 |
| France | 50,059 | 46,206 |
| Germany | 57,344 | 53,866 |
| Greece | 24,268 | 28,733 |
| India | 10,805 | 50,517 |
| Italy | 35,267 | 32,705 |
| Israel | 42,862 | 58,826 |
| Japan | 70,927 | 64,240 |
| South Korea | 49,336 | 42,876 |
| Mexico | 27,356 | 25,108 |
| Netherlands | 56,750 | 53,542 |
| Norway | 56,136 | 38,257 |
| Poland | 11,531 | 20,883 |
| Portugal | 27,386 | 30,860 |
| Russia | 10,873 | 18,006 |
| Spain | 32,078 | 35,102 |
| Sweden | 57,271 | 47,334 |
| Switzerland | 81,776 | 58,948 |
| Turkey | 18,867 | 27,001 |
| United Kingdom | 56,181 | 51,880 |
| United States | 59,330 | 60,725 |
| Worldwide (weighted average) | 48,701 | 47,837 |

Source: Authors' elaborations based on Morretta et al. (2022). New elaborations refer to years from 1990 to 1997 and from 2019 to 2021 and to the following countries: Brazil, Canada, South Korea, Mexico, Russia. Values in PPS have been obtained by using the World Bank's PPP conversion factor. Weights entering the worldwide weighted average are those in Figure 10.

We randomly cross-checked our salary data with proprietary and national data. For instance, and considering the first two countries with the largest share of LHC scientists, we cross-check the 2021 salary for the USA in our database (EUR 60,156 in Appendix E) with the average US PayScale[29] data for different positions in academia (assistant professor, professor, assistant researcher, fellow, etc.) resulting in an average salary of EUR 65,316. The same exercise for Germany resulted in an average salary in our database of (EUR 68,768 in Appendix E) against an average salary of EUR 67,682, combining German national sources and Glassdoor data.[30] When relevant discrepancies existed, we relied on Moretta et al. (2022) data.[31]

## 5.4 Time dedicated to research

The work of researchers in academia is often split between research, teaching, administrative tasks, and service responsibilities (e.g., committee work, peer reviewing) (OECD, 2016). They may spend between 20% and 80% of their total working time on writing and preparing scientific publications, though this can vary significantly based on personal factors, institutional expectations, and disciplinary requirements (Teichler et al., 2018; Bentley and Kyvik, 2012; Bazeley, 2010). Early-career physicists (graduate students and postdocs) spend 70-90% of their time on research. Assistant professors are typically expected to develop their own research programs while also taking on teaching and service responsibilities. They may spend around 50-70% of their time on research, especially in the early years before securing tenure. Associate and full professors (mid-career and senior faculty) generally spend around 30-60% of their time on research but may vary significantly based on whether they take on additional roles (like department chair or program director) that reduce their research time.

The institution type also affects the time dedicated to research. Research-intensive universities (e.g., R1 and R2 institutions in the USA), especially in STEM, often approach 70% to 80%, with a significant portion of that time dedicated to writing and publishing. Conversely, at teaching-focused institutions (e.g. liberal arts colleges), research time may be reduced to 20% to 40%, with the rest of their time spent on teaching and other duties (Ylijok and Ursin, 2021; Goodwin and Deneef; 2020; Teichler et al., 2018; Nature, 2016). At CERN, researchers often dedicate between 50% and 80% of their time to conducting experiments, analysing data, and preparing scientific publications (FCCIS, 2024).

Based on this evidence and the existing literature on the evaluation of scientific publications, we set the parameter $h_{res}$ at 65%. This value represents a midpoint estimate that reflects a balance between academic expectations and the intended research demand in research infrastructures (Morretta et al., 2022; Florio et al., 2016; 2022).[32] Since we ignore the true value of this parameter (unless a dedicated study is done), we apply a sensitivity analysis by assuming different values the parameter $h_{res}$ and examine how our results respond to these changes (Section 6.3).

---

[29] The American PayScale database is a comprehensive resource that provides detailed information on salaries, compensation packages, and employment trends across various industries and job roles in the United States and in other few countries. See https://www.payscale.com/en-eu/
[30] For German national sources see https://www.salaryexpert.com/salary/job/associate-professor/germany . Last access in October 2024. As regards, Glassdoor, we considered the same academic positions as in the USA. Glassdoor is a popular online platform that provides insights into company culture, salaries, and job opportunities. The database includes, among others, Saraly information, company reviews, job listings. See https://www.glassdoor.com/
[31] For instance, in the case of Switzerland, our database indicates a salary of EUR 82,725 against a salary of EUR 104, 000 in PayScale.

[32] The value of 65% was also used in similar studies in HEP (Florio er al. 2016; Bastianin and Florio, 2018: Bastianin et al., 2022).

## 5.5 The marginal cost of publishing

The marginal cost of publishing ($MPC_{pub_t}$) is the final element needed to assess the marginal cost of publication in our framework. We treat it as an exogenous variable in Eq. 3 because it is not the primary focus of our analysis, and its value is relatively marginal compared to the cost of conducting research and drafting a scientific product (see also Section 6.3). Additionally, as outlined in Section 4, HEP researchers have long explored alternative methods of disseminating scientific findings, moving away from traditional peer-reviewed journals. Morretta al. (2022) conducted an extensive literature review on the cost of publishing across different types of journals and editors. They analysis provided time series data on publishing costs from 1998 to 2018 (Morretta et al., 2022: Table 9), distinguishing between journal articles and books, which typically go through a comprehensive pee-review and editorial process (from submission to peer-review, publication, indexing and archiving) and other types of scientific products, including conferences proceeding, which instead follow a different, less demanding publishing path. As with the salary data, we followed the same methodology as Morretta et al. (2022) to extend the time coverage for the period from 1990 to 2021 with the results visualised in Table 4.

**Table 4. The marginal costs of publishing for publishers, by year (1990 -2021).**

| Year | Conference proceedings | | Journal articles and books | | Average all scientific products | |
|---|---|---|---|---|---|---|
| | EUR | PPS | EUR | PPS | EUR | PPS |
| 1990 | 163 | 129 | 498 | 393 | 331 | 261 |
| 1991 | 165 | 130 | 502 | 397 | 334 | 264 |
| 1992 | 166 | 132 | 507 | 401 | 337 | 266 |
| 1993 | 168 | 133 | 512 | 405 | 340 | 269 |
| 1994 | 169 | 134 | 516 | 409 | 343 | 272 |
| 1995 | 171 | 136 | 521 | 413 | 346 | 274 |
| 1996 | 173 | 137 | 526 | 417 | 349 | 277 |
| 1997 | 174 | 138 | 531 | 421 | 352 | 280 |
| 1998 | 176 | 140 | 536 | 425 | 356 | 282 |
| 1999 | 178 | 142 | 543 | 431 | 361 | 286 |
| 2000 | 182 | 145 | 555 | 441 | 369 | 293 |
| 2001 | 190 | 147 | 580 | 449 | 385 | 298 |
| 2002 | 183 | 148 | 558 | 452 | 371 | 300 |
| 2003 | 157 | 152 | 478 | 464 | 318 | 308 |
| 2004 | 147 | 156 | 447 | 476 | 297 | 316 |
| 2005 | 151 | 160 | 461 | 488 | 306 | 324 |
| 2006 | 154 | 160 | 468 | 487 | 311 | 324 |

| Year | | | | | | |
|------|-----|-----|-----|-----|-----|-----|
| 2007 | 145 | 163 | 442 | 497 | 293 | 330 |
| 2008 | 138 | 163 | 420 | 497 | 279 | 330 |
| 2009 | 147 | 162 | 447 | 494 | 297 | 328 |
| 2010 | 155 | 164 | 473 | 498 | 314 | 331 |
| 2011 | 152 | 164 | 462 | 501 | 307 | 333 |
| 2012 | 167 | 166 | 507 | 507 | 337 | 336 |
| 2013 | 164 | 165 | 501 | 503 | 333 | 334 |
| 2014 | 167 | 167 | 510 | 510 | 339 | 339 |
| 2015 | 203 | 170 | 618 | 518 | 410 | 344 |
| 2016 | 205 | 166 | 625 | 506 | 415 | 336 |
| 2017 | 205 | 167 | 625 | 509 | 415 | 338 |
| 2018 | 201 | 169 | 613 | 516 | 407 | 342 |
| 2019 | 203 | 171 | 618 | 521 | 411 | 346 |
| 2020 | 205 | 173 | 624 | 526 | 414 | 349 |
| 2021 | 207 | 174 | 630 | 531 | 418 | 353 |
| Average cost | 173 | 154 | 527 | 469 | 350 | 311 |

Source: *Authors' elaborations based on Morretta et al. (2022). New elaborations refer to years from 1990 to 1997 and from 2019 to 2021. Value in PPS have been obtained by using the World Bank's global PPP conversion factor.*

# 6. Results

## 6.1 The monetary value

### 6.1.1 All scientific products

In a first step, we considered the entire universe of scientific products in our database, including both non-related and related-LHC products. We estimated the monetary value of these products by applying the MPC formula in Eq. 3 and setting the parameters as follows (sub-scripts omitted for simplicity): the productivity function $y_{res_t}(n_t)$ obtained empirically in **Figure 9** (Section 5.2), the average scientists' salary ($w_{res}$) at EUR 47,837 PPS (Section 5.3), the average time dedicated to research ($h_{res}$) at 65% (Section 5.4), and the average marginal cost of publishing ($MPC_{pub}$) at EUR 311 PPS (Section 5.5). For the sake of clarity, we only allowed the productivity function to vary over time and by the number of co-authors. We kept all the other parameters fixed at their averages because we wanted to focus on how the productivity function shapes the monetary values of scientific products, removing all the other sources of variability. Our conclusions remain valid if the other parameters are left to vary over time as well. **Figure 11** shows that the monetary values range from a minimum of EUR PPS 10,078 for single-authored products to a maximum of EUR PPS 1,564,393 for highly collaborative products, with a co-authorship involving 3,000 co-authors. The distribution is right-skewed, with the mean slightly above the third quantile (Q75). The mean value is EUR PPS 84,210, significantly higher than the median (EUR PPS 38,222), reflecting the presence of high-value collaborative products, particularly LHC products, as discussed below.

**Figure 11**. **Distribution of the monetary values of all scientific products by size of co-authorship (1990 - 2021).**

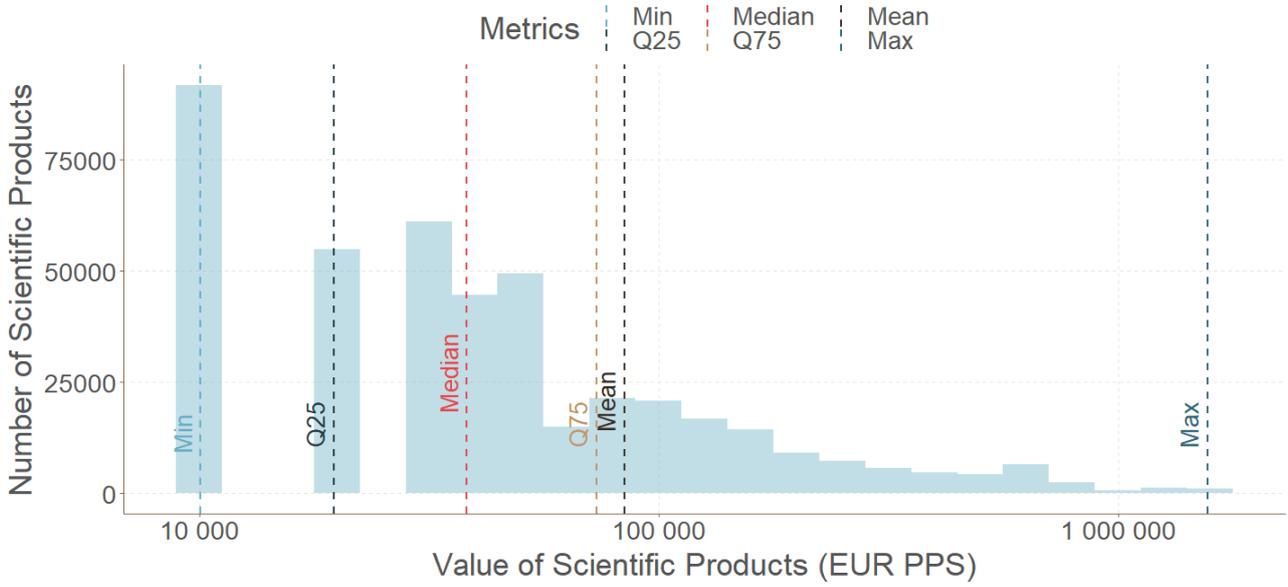

Source: Authors based on Eq. 3. The sample consists of 433,903 scientific products trimmed at 3,000 co-authors. it includes all the products of identifiable authors with at least one LHC-related product in 1990-2021, i.e., both their LHC-related and non-LHC related products. Apart from the productivity function, the value of the other parameters entering Eq. 3 were set as at their average values: $\overline{MPC}_{pub} = 311$; $\bar{h}_{res}= 65\%$, $\overline{w}_{res} = 47,837$. We also run simulations with varying values of $\overline{MPC}_{pub}$, $\overline{w}_{res}$, and $\bar{h}_{res}$ over time (see sensitivity analysis in Section 6.3).

### 6.1.2 LHC-related scientific products

In Section 5, we showed that the distribution of the LHC scientific products considerably differs from the distribution of the total scientific production by co-authorship and type of products. So, one might be interested in the monetary value of the LHC products (**Figure 12).** As expected, the distribution is right-skewed, with the median (and Q25) close to the minimum value (EUR PPS 10,078), largely driven by single-authored products such as doctoral theses, notes, and conference proceedings, which account for more than 80% of LHC-related scientific products in our sample (Section 5.1). The mean value is EUR PPS 143,150, reflecting the higher share of products authored by the LHC collaborations compared to the entire universe of products. Consistently with the fact that LHC-related products represent a sub-set of the entire universe, the minimum and the maximum value (EUR PPS of 1,564,393) overlap with the extremes already observed for the universe (see also Appendix C on this point).

**Figure 12**. Distribution of the monetary values of LHC-related scientific products by size of co-authorship (1990 - 2021).

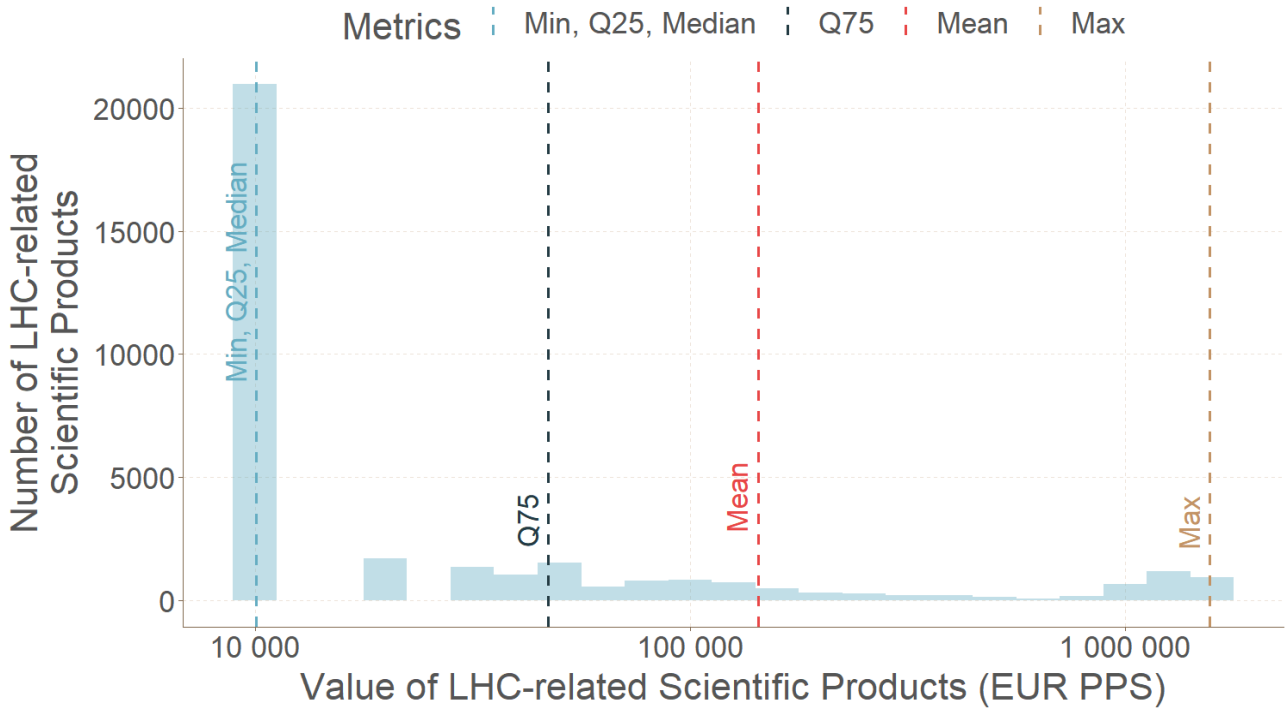

Source: Authors based on Eq. 3 applied to the empirical distribution of LHC-related scientific products. The sample consists of 34,397 LHC-related scientific products trimmed at 3,000 co-authors (1990-2021). The value of the parameters entering Eq. 3 were set as at their average values: $\overline{MPC}_{pub} = 311; \bar{h}_{res} = 65\%, \overline{w}_{res} = 47,837$.

Articles and pre-prints better reflect the progress and contributions of collaborative research at LHC compared to other types of scientific products. The distribution of their monetary values is illustrated in **Figure 13** which tells us a different story from what discussed earlier. 50% of the articles and pre-prints have a monetary value above EUR PPS 966,896 per product with a mean value of EUR PPS 744,472. These high values correspond to the extensive collaborative efforts required for groundbreaking research at the LHC, often involving hundreds (if not thousands) of scientists. Additionally, these findings are consistent with the observation that research articles typically reflect higher quality and exert greater influence on the scientific community, as will be discussed in the next section.

**Figure 13. Distribution of the monetary values of LHC-related articles and pre-prints by size of co-authorship (1990 - 2021)**

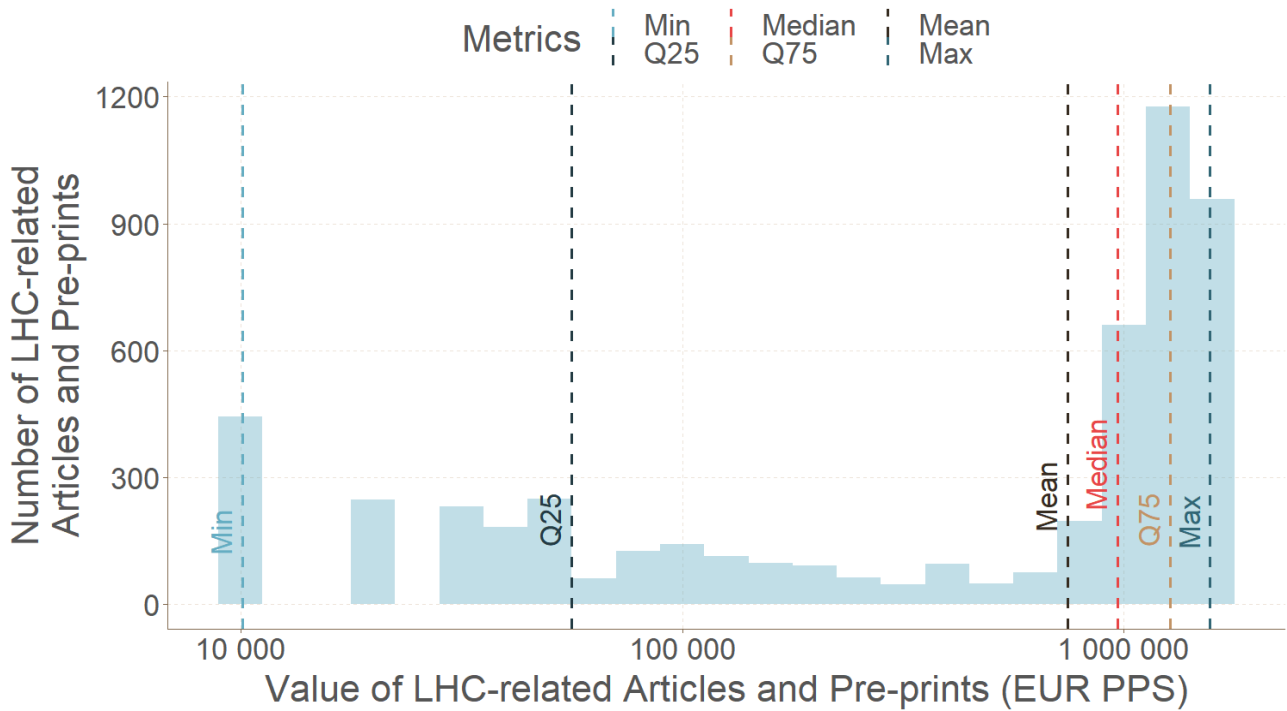

Source: Authors based on Eq. 3 applied to the empirical distribution of LHC-related articles and pre-prints. The sample consists of 5,292 LHC-related articles trimmed at 3,000 co-authors (1990-2021). The value of the parameters entering Eq. 3 were set as at their average values: $\overline{MPC}_{pub} = 311$; $\bar{h}_{res} = 65\%$, $\overline{w}_{res} = 47{,}837$.

## 6.2 Higher monetary value and quality

A key implication of our findings is that, on average, a greater number of co-authors correlates with a higher monetary value of the scientific outputs, suggesting that collaborative papers tend to be of higher quality. Numerous studies have explored the relationship between co-authorship and scientific excellence (Thelwall et al., 2023; Isfandyari-Moghaddam et al., 2023; Abbasi et al., 2011). Multi-authored papers frequently benefit from the contributions of researchers across different areas of expertise, often resulting in more comprehensive and robust studies. This interdisciplinary collaboration enhances the quality of research, as diverse perspectives contribute to greater rigour, depth, and scope (Abramo et al., 2017a; Larivière et al., 2015a). In disciplines such as physics - including high-energy physics, as in our case - medicine, and biology, large-scale experiments and complex data analyses are common. Multiple authorships are thus the norm and are often essential for producing high-quality research. Furthermore, the networking, international collaboration, and strategic partnerships inherent in large-scale scientific projects like the LHC facilitate the exchange of ideas, knowledge spillover, and access to greater resources, all of which contribute to the creation of new knowledge and the production of higher-quality work (Jang and Ko, 2019; Manganote et al., 2016). To test this hypothesis, we use citations as a proxy for the impact or visibility of research (Yadav et al., 2023).

We retrieved the number of citations obtained by all the scientific products in our sample (LHC-related and non-LHC-related) from 1990 to 2021.[33] We acknowledged that the "publish or perish" phenomenon in academia (Yadav et al., 2023; Kirkpatrick et al., 2022; Furnham, 2021) and (self-)citations inflation in multi-authored products may distort results. On the other hand, self-citations are inevitable in "closed" scientific fields such as HEP. Accordingly, we preset citation data with and without self-citations (**Table** ).[34] Overall, more than 11.9 million citations were received by the scientific products in the period considered, with research articles and pre-prints receiving 93% of all citations (**Table 5, Panel A**).[35] Excluding self-citations resulted in fewer overall but did not diminish the significance of articles, underscoring their undeniable impact on the global HEP scientific community compared to other types of scientific products (**Table 5, Panel B**).

**Table 53. Number of citations received by all scientific products (1990 -2021).**

| Scientific product | Total (%) | Mean | Std. Dev | Min | Max |
|---|---|---|---|---|---|
| PANEL A: all citations | | | | | |
| Conference papers and proceeding | 733,230 (6%) | 4.2 | 27.8 | 0 | 4,437 |
| **Articles and pre-prints** | **11,071,660 (93%)** | **47.3** | **178.1** | **0** | **19,043** |
| Theses | 12,694 (0.1%) | 1.0 | 15.4 | 0 | 1,487 |
| Notes | 16,436 (0.1%) | 2.2 | 25,7 | 0 | 2,033 |
| Books and books chapters | 35,568 (0.3%) | 21.1 | 142.4 | 0 | 4,879 |
| Reports | 56,697 (0.5%) | 186.5 | 485.4 | 0 | 4,726 |
| All scientific products | 11,926,285 (100%) | 27.6 | 135.2 | 0 | 19,043 |
| PANEL B: excluding self-citations | | | | | |
| Conference papers and proceeding | 517,517 (5.7%) | 2.9 | 24.0 | 0 | 4,112 |
| **Articles and pre-prints** | **8,495,267 (93.3%)** | **36.3** | **157.8** | **0** | **16,320** |
| Theses | 10,797 (0.1%) | 0.9 | 14.8 | 0 | 1,446 |
| Notes | 12,007 (0.1%) | 1.6 | 20.1 | 0 | 1,593 |
| Book and books chapters | 31,317 (0.3%) | 18.6 | 137.7 | 0 | 4,838 |
| Reports | 38,162 (0.4%) | 125.5 | 368.4 | 0 | 4,608 |
| All scientific products | 9,105,067 | 21.1 | 119.2 | 0 | 16,320 |

Source: Authors' elaborations based on INSPIRE data for 431,773 scientific products involving the LHC authors (1990-2021) with maximum 3,000 authors. It includes both LHC-related and non-LHC-related products. Citations can include those received outside this timeframe. Self-citations exclude citations by the authors of the publication, or by the same collaboration, as per INSPIRE guidelines.

The analysis of the LHC-related scientific products is consistent with the analysis of the universe of all products, with LHC research articles and pre-prints receiving the lion's share of citations of all the LHC scientific products, either including or excluding self-citations (**Table 6**).

---

[33] It may include some citations received at a later stage, e.g., from 2022 onwards. Scientific products with more than 3,000 authors were excluded.
[34] Self-citations exclude the citations produced by any author or a product, or from the same collaboration. See https://blog.inspirehep.net/2020/06/whats-new-self-citations-author-citations-and-hepdata-links/
[35] As discussed in Section 5.1, experimental HEP proceedings material is only partially original material because a good part will be found again in a research paper at some stage. That is consistent with the fact that proceedings material is much less cited than research articles and pre-prints.

**Table 6. Number of citations received by the LHC-related scientific products (1990 -2021).**

| Scientific product | Total | Mean | Std. Dev | Min | Max |
|---|---:|---:|---:|---:|---:|
| PANEL A: all citations | | | | | |
| Conference papers and proceeding | 50,765 (9%) | 2.7 | 12.1 | 0 | 926 |
| **Article** | **486,479 (85%)** | **91.9** | **371.7** | **0** | **15,505** |
| Theses | 2,701 (0.5%) | 0.5 | 1.7 | 0 | 55 |
| Notes | 11,584 (2%) | 3.0 | 12.1 | 0 | 355 |
| Books and books chapters | 543 (0.1%) | 10.2 | 39.6 | 0 | 268 |
| Reports | 21,188 (3.7%) | 450.8 | 814.8 | 1 | 4,726 |
| All scientific products | 573,260 (100%) | 17.2 | 155.7 | 0 | 15,505 |
| PANEL B: excluding self-citations | | | | | |
| Conference papers and proceeding | 35,517 (11%) | 1.9 | 7.7 | 0 | 359 |
| **Articles** | **259,970 (81.3%)** | **49.1** | **258.3** | **0** | **12,444** |
| Theses | 2,283 (0.7%) | 0.4 | 1.4 | 0 | 35 |
| Notes | 8,677 (2.7%) | 2.2 | 9.7 | 0 | 317 |
| Book and books chapters | 348 (0.1%) | 6.6 | 23.6 | 0 | 147 |
| Reports | 12,828 (4.0%) | 272.9 | 466.0 | 0 | 2,592 |
| All scientific products | 319,623 (100%) | 9.6 | 106.4 | 0 | 12,444 |

Source: Authors' elaborations based on INSPIRE data for 33,573 LHC-related scientific products (1990-2021) with maximum 3,000 authors. Citations can include those received outside this timeframe. Self-citations exclude citations by the authors of the publication, or by the same collaboration, as per INSPIRE guidelines.

To further investigate the topic, **Figure 14** shows the correlation between the number of citations and co-authorship for all the scientific products of the sample (LHC-related and non-LHC-related). The left panel includes all citations, while the right panel excludes self-citations. In both cases, there is a weak positive but statistically significant relationship between these variables at 1%, with correlation coefficients of 0.1 (*p-value < 2.2e-16*) and 0.06 (*p-value < 2.2e-16*), respectively. The correlation coefficient is higher for the LHC-related scientific products, namely 0.3 if considering all the citations and 0.2 if excluding self-citations (*p-value < 2.2e-16* statistically significant at more than 1% level) (**Figure 15**). These findings suggest that co-authorship is associated with a higher number of citations, indicating more impactful and higher-quality scientific activity. This aligns with the existing literature on this topic. Research articles and reports are the primary drivers of this relationship.

**Figure 14. Number of citations received by all scientific products by size of co-authorship (1990-2021).**

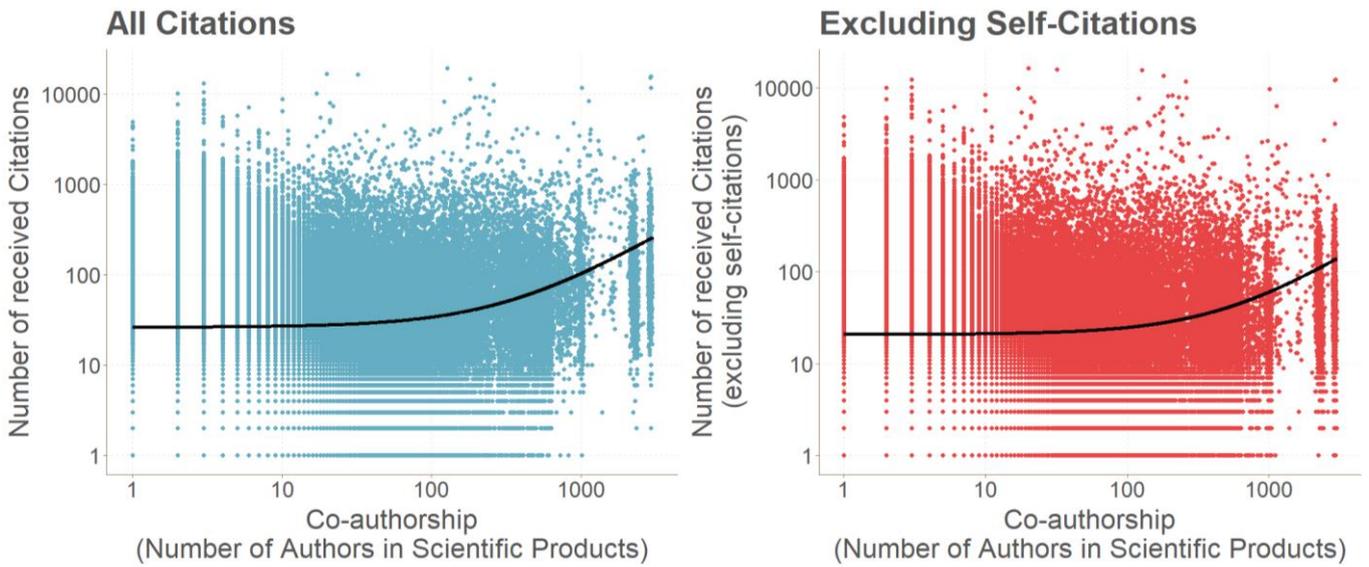

Source: Full sample of 433 903 scientific products with less than 3000 authors (1990-2021). It includes both LHC-related and non-LHC-related scientific products. Each point represents a unique scientific product of this sample. Citations and Citations excluding self-citations were modified by adding one to them to allow log scales for better visibility. Black curves are linear regressions.

**Figure 15. Number of citations received by LHC-related scientific products by size of co-authorship (1990-2021).**

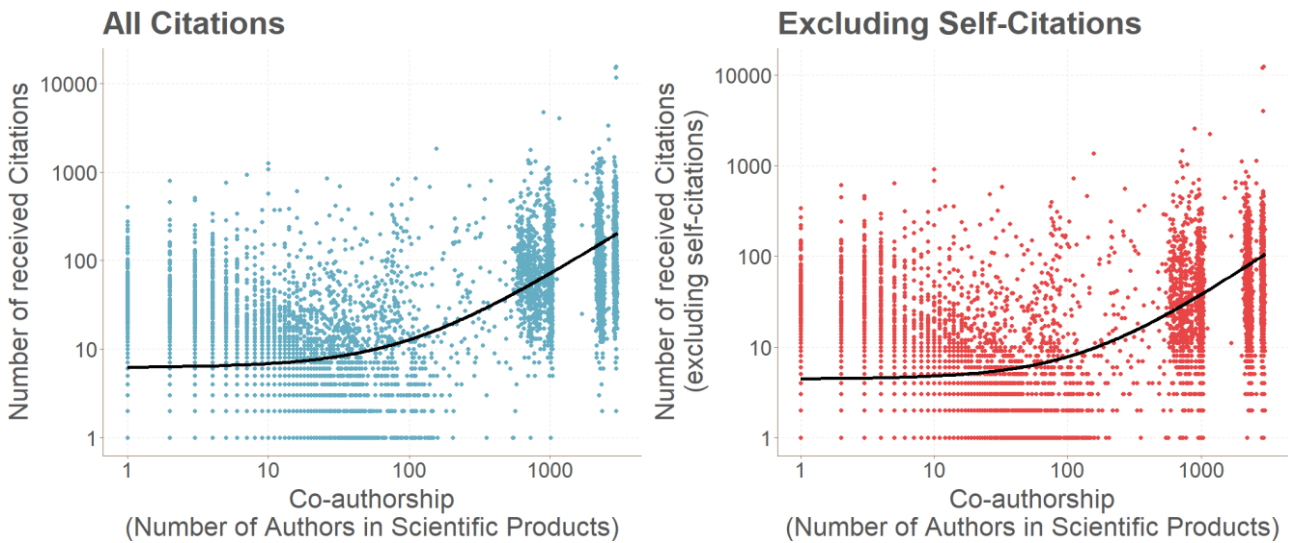

Source: Full sample of 34 397 LHC-related scientific products with less than 3000 authors (1990-2021). Each point represents a unique LHC-related scientific product of this sample. Citations and Citations excluding self-citations were modified by adding one to them to allow log scales for better visibility. Black curves are linear regressions.

### 6.3 Sensitivity analysis

Morretta et al. (2022) suggest that a future avenue of research could involve conducting sensitivity analyses to assess how the monetary values of scientific products fluctuate based on key model parameters, thus testing the robustness of the assumptions. Similarly, Rosseau et al. (2021) propose exploring how factors like the number of co-authors affect the perceived value or evaluation of publications.

**Table 7** illustrates the impact of varying model parameters in Eq. 3 on the monetary value of scientific products across different scenarios, with 1%, 5% and 10% increases (percentage points in the case of $h_{res}$, which is measured in percentage). The number of co-authors ($n$) in Eq.3 enters the equation both at the denominator (via the productivity function, with an expected negative), and at the numerator (accounting for co-authorship, with an expected positive impact). A 1% increase in the number of co-authors results in a less than 1% increase in the monetary value of scientific products, indicating that "the co-authorship effect" outweighs the "productivity effect". The interpretation is straightforward. According to the MPC approach, while co-authorship increases the contributions to a product, it may not proportionally enhance individual productivity, as shown by the non-linearity of the productivity-co-authorship relationship in Section 5.2[36]

As expected, salary and the time dedicated to research are other important parameters. The monetary value of scientific products is highly sensitive to these, with (more than) proportional impacts on the monetary value. By contrast, the marginal cost of publishing has a negligible impact, close to zero. Table E.3 in Appendix E shows the same analysis by simulating decreasing scenarios (-1%, -5%, and -10%), leading to the same results and interpretations.

**Table 7. Impact of the parameters on the monetary value of scientific products based on different scenarios (holding other variables constant)**

| Variables | Increase compared to baseline value | | |
|---|---|---|---|
| | +1% | +5% | +10% |
| $n$ (number of co-authors) | +0.3% | +2.5% | +6.5% |
| $W_{res}$ (salary) | +1% | +5% | +10% |
| $h_{res}$ (time dedicated to research) | +1.5% | +7.7% | +15.3% |
| $MCP_{pub}$ (marginal cost of publishing) | +0% | +0.02% | +0.04% |

Source: Authors simulation based on the mentioned deviations from the standard parameters. For the number of authors, the entire distribution is shifted uniformly (e.g., publications with 10 authors increase to 11 authors if the number is scaled by 10%, publications with 100 authors to 110). The number of authors is rounded. Parameters' subscripts omitted for simplicity.

## 7. Discussion

In recent years, there has been growing interest in monetarily valuing scientific publications (Morretta et al., 2022; Florio, 2019; Florio et al., 2016; Battistoni et al., 2016). While previous studies represent progress, they fail to adequately account for the increasing prevalence of collaboration in research, disregarding the widespread advocacy from policymakers, scientific communities, and businesses for inter-organisational and international scientific collaboration to address global challenges (Abbass et al., 2022; Jessel et al., 2019). Additionally, these studies do not effectively assess scientists' productivity over time, leading to an incomplete understanding of their contributions to their respective fields. Our study contributes to this literature by examining the relationship between scientific collaboration and productivity, corroborating previous, mostly anecdotal and qualitative insights on this topic.

---

[36] The interpretation holds true on average. Since the productivity function is not linear, the sensitivity of the MPC to a variation of the number of authors depends on the starting point of the productivity function.

We developed a generalised framework based on the marginal cost approach and applied it to the HEP field at CERN's LHC. This allowed us to explore how collaboration (co-authorship) influences the monetary value of scientific publications, specifically considering its effect on individual scientist's productivity. The value of scientific products varies significantly depending on the size of the collaborations and the type of product. In HEP (all scientific products), for example, the monetary value ranges from around EUR PPS 10,000 for single-authored products to EUR PPS 1.56 million for research articles with approximately 3,000 co-authors. Research articles tend to have a higher value than other products, likely because they present original findings and discoveries. Furthermore, our research includes a citation analysis (with and without self-citations), suggesting that research collaborations have an intrinsically higher value (quality) due to various factors beyond the authors' networks. Collaborative papers at the LHC benefit from the diverse expertise of authors from different subfields, often leading to more comprehensive and robust research. CERN scientific products, especially those from major LHC experiments, are highly cited in the scientific community due to their significant contribution to advancing our understanding of fundamental physics. The implications of our research are both methodological and policy oriented.

Methodologically, it provides a framework for assessing collaboration in research and rethinking how monetary value can be attributed to scientific publications by integrating the link between collaboration and scientists' productivity. Failing to adequately account for these factors, as seen in previous studies (Bastianin et al., 2022; Bastianin and Florio, 2018; Florio et al., 2016; Battistoni et al., 2016) leads to severe downward biases in the estimation of the value of scientific products, caused by skewed distributions and arbitrary assumptions. This result would be an underestimation of the social value of scientific publications.

Second, we argue that summary statistics, such as average or median monetary values, are inadequate for capturing the complexity of the knowledge spillovers generated by research collaboration and their impact on the monetary value of scientific products. To produce credible results, it is essential to consider the entire distribution of values, as shown by the variation in co-authorship sizes (as in **Figures 10 – 13**). Additionally, sensitivity analyses of key parameters in the model should be incorporated to understand their influence on the results fully.

Third, our research framework can be easily applied to fields beyond physics where scientific collaboration is vital, including research in genomics, life sciences, health, and climate science, among others, and to European-funded research programmes such as Horizon Europe. We used R, an open-source software, to extract and process all the data following the procedure detailed in Appendix D. The data we used are freely accessible through online bibliographic databases, consisting of around half a million scientific products that include detailed information on co-authors, as outlined in Section 5. The data should be primarily used to set the productivity parameter according to the size of the collaboration and practices in the specific field under analysis. Then, data on scientists' salaries and the proportion of time dedicated to the research can be incorporated.

From a science policy viewpoint, our findings highlight the critical role of scientific collaboration as a mechanism for producing high-quality and high-value scientific outputs. Large, international collaborations, particularly those involving researchers across multiple disciplines, are essential for addressing modern societal challenges. These efforts help generate solutions that would not emerge from individual scientists, a single discipline alone or single countries. Decision-makers can recognise the value in fostering wide research collaboration endeavours and these intersections and strategically direct resources toward research that tackles complex, interdisciplinary problems with high potential payoffs. Collaborative research accelerates the pace of discovery by enabling shared resources and knowledge, which in turn reduces time-to-impact. This rapid advancement benefits policymakers by identifying breakthroughs early, allowing for timely investments in areas poised to produce transformative results. The example of CERN and its LHC scientific programme in this study showcases the importance of collaboration and coordination when developing large R&I infrastructure projects (Draghi report, 2024) and can be seen as supra-national hyperscale projects that can serve as a benchmark for big science investments. Overall, supporting scientific collaboration not only aligns research with policy needs but also strengthens the link between science and policy, leading to more informed decision-making and more impactful research outcomes.

This study has a number of limitations.

First, we focus on the LHC research collaborations/experiments and implicitly assume that co-authorship (used as a proxy of collaboration) fully captures the extent of collaboration. However, not all LHC research collaborations result in co-authored scientific products (Aksnes and Sivertsen, 2023). In experimental collaborations, articles often feature a long list of authors, some of whom may not have directly contributed to the writing. In contrast, some articles are signed only by the writers, even though many others played a role in the experimental work. Therefore, co-authorship may not perfectly reflect the full scope of LHC collaboration and could be a partial indicator. To give an example, the ATLAS experiment is one of the largest collaborative efforts in science, with approximately 6,000 members (physicists, engineers, technicians, students and support staff from around the world) and 3,000 scientific authors (mainly physicists and engineers).[37] In our study, we analysed scientific products with up to 5,246 authors, and used a truncated working sub-sample up to a maximum 3,000 when calculating productivity and monetary value. While this truncated sample aligns with the number of ATLAS physicists and engineers, providing some reassurance regarding the robustness of our findings, it is still possible that we did not entirely capture the full collaboration efforts.

Second, our estimation of the scientists' productivity assumes that all the authors in our sample are "active" even if they have not produced any scientific product in a given year, whether LHC-related or not. For instance, this may occur due to temporary career interruptions, delays in the peer-review process, or other reasons that we cannot observe in the data. This assumption could lead to erroneous attributions in specific cases, such as career changes, unobserved documents in the INSPIRE HEP, or other edge cases. However, excluding scientists with zero production in a given year would result in an inflated and biased estimate of productivity. Future research could address this limitation by reconstructing the full career trajectories of the involved researchers, which would require extensive data collection efforts. This level of detail might not only be feasible for a sub-sample of authors.

---

[37] See ATLAS experiment web page at the following url. https://atlas.cern/Discover/Collaboration#:~:text=Organisational%20Structure,members%20and%203000%20scientific%20authors . Last access in October, 2024.

Third, one could argue that accurately estimating a scientist's productivity requires identifying individual contributions to the scientific products, with the first-listed author typically carrying more weight. (such as in life sciences publications). While some attempts have been made (Flores-Szwagrzak and Treibich, 2015), assigning such weights is challenging, as it is notoriously difficult to develop an objective and comparable quality measure for publications across different disciplines (Rosseau et al., 2021; Wilsdon et al., 2015). In the case of the LHC, this approach would conflict with its egalitarian tradition of listing authors alphabetically, which is intended to acknowledge the collaborative nature of the work.

Fourth, we document those large collaborations, via the productivity function, not only increase research efficiency (allowing authors to produce more scientific products in a given period) but are also associated with higher quality (and monetary value) because of the division of labour, knowledge and skills sharing, and improved access to resources and equipment. In our framework, one might expect a one-to-one correspondence between the number of co-authors and the monetary value of scientific products. While this is true on average, the productivity function is neither linear nor injective, meaning that a specific monetary value (a particular element in the codomain) can be associated with different levels of co-authorship (elements in the domain). Figure E.1 in Appendix E presents a simulation of the unitary monetary value of a generic scientific product, where the relationship between co-authorship and value is fully determined by the productivity function, which is empirically observed. In other words, scientific collaborations tend to yield increasingly economically valuable products, though the exact process, the "technology", and the underlying mechanisms driving this production are not fully understood. Effective research collaboration involves more than merely adding names to a paper; it requires trust, respect, and a shared commitment to achieving mutual goals. In a counterfactual scenario where authors did not collaborate, the productivity function would likely yield different insights regarding the volume, type, and quality of publications produced over time. Further research is welcomed to better model the productivity function.[38]

Lastly, the positive relationship between multi-authored scientific products and research quality, especially in research articles, raises the question of whether the marginal cost approach is sufficient to fully capture the value of the research articles that contain original or breakthrough ideas. While the MPC estimates the cost of producing publications, it does not account for the intrinsic value of the ideas or scientific knowledge embedded in the paper (see Section 2). As Morretta et al. (2022:13) point out, the MPC approach "*does not include any social use of the knowledge embodied in the papers*". Although this topic lies beyond the scope of our research, we agree with Morretta et al. (2022) that it warrants further exploration. We also support the suggestions by Rousseau et al. (2021), who propose using survey-based approaches, such as contingent valuation methods or choice experiment modelling, to examine both the use and non-use values of research as complementary evidence to MPC values. These surveys could target a representative sample of the scientific community, regardless of location or size, to elicit and infer the value of publications based on the knowledge they incorporate, their potential for breakthrough discoveries, and other key factors. The challenge lies in designing effective questions to gauge the value of scientific publications, which could include scenario analyses informed by the scientific community. Surveys can also be used to infer values of the parameters related to the research time ($h_{res}$ or $\alpha$-like parameters). How authors allocate their time among their catalogue of scientific products is not known a priori. Multi-country surveys to assess non-use values already exist and could be adapted to value scientific publications from global collaborations (Secci et al., 2023; Rousseau et al., 2021). After all, the primary mission of research, especially large-scale research infrastructures, is to generate new insights, theories, and breakthrough discoveries in their fields. This contribution must be evaluated to some extent. If, as Florio et al. (2016) argue, the implicit price the scientific community is willing to pay for an additional publication is at least equal to its production cost, then the estimation of 'demand-side' monetary values through survey-based methods can offer empirical validation.

---

[38] For instance, further research can help explicitly integrating into the model quality metrics, such as, among others, the number of received citations of specific products. One might be tempted to integrate citations as an additive or multiplicative factor of the entire MPC formula. This would create overlaps because quality is already somewhat captured by the productivity function. The challenge is to develop a theoretical refined model of the publication production process, including quality.

# ANNEX
# Appendix A. The derivation of the MPC formula

The business model in scientific research publishing (SQW, 2004) indicates that the cost function of peer-reviewed scientific publications at the time $t$ comprises two main components: i) the cost of doing research and writing the paper ($Cost_{res}$), and ii) the cost of publishing it ($Cost_{pub}$) (Eq. A.1). For the sake of simplicity, we omit the subscript $t$ in this appendix.

$$Total\ Cost_{scpub} = f\ (Cost_{res}) + f(Cost_{pub}) \qquad A.2$$

Following the same denotation of Morretta et al. (2022) for the sake of comparability, and expanding eq. A.1 to visualise the cost components, the social total cost to produce scientific publications is given by Eq. A.2:

$$Total\ Cost_{scpub} = f\ (K_{res}, L_{res}, L_{res}', OP_{res},) + \gamma\ (K_{pub}, L_{pub}, L_{pub}', OP_{pub}) \qquad A.2$$

Where $K_{res}$ represents the capital expenditure required to create publishable work, primarily funded by the research institution to which the scientist or author is affiliated. This includes the expenses for equipment, computing resources, software, databases, and other fixed assets essential for research. $L_{res}$ is the cost of labour costs of scientists, proxied by the gross salary. $L_{res}'$ is the labour cost of the administrative and technical staff, often not directly involved in research activity, but necessary to support it; and , $OP_{res}$, denotes other operating costs supported by the research institutes, including costs of the trips for conferences, electricity, etc.).

Similarly, the cost of publishing is made of the capital expenditure of publishers ($K_{pub}$,), including any fixed asset to create online and printed publication); $L_{pub}$ denotes the labour cost of publishers involved in editing and the peer-review process; $L_{pub}'$ is the labour cost of the administrative and technical necessary to publish (e.g. proofreaders, copywriters etc.),, and finally $OP_{pub}$ covers all other publisher operating costs (e.g., dissemination of printed journals and maintenance of digital platforms for online journals).

In Morretta et al. (2022), Eq. A.2 is augmented by externalities to take on board external effects of the research process (e.g., use of toxic reagents, ethical issues, etc.) and publication process (e.g., $CO_2$ emission for printing and distributing pieces of paper). Taking into account externalities is necessary in a Cost-Benefit Analysis (CBA) framework as the main goal is to estimate the social value of publications, including non-market effects (Florio, 2019). While useful, including externalities in this study is not necessary and neglecting them does not change the main results.

Following SQW (2004), the total marginal production cost (MPC) of scientific publications is given by differentiating Eq. A.2 with respect to the number of publications. (Eq. A.3). In this process, it can be assumed that all capital expenditure, the labour cost of the administrative and technical staff, and operating costs of research and publication are fixed, in the sense that do not depend (or only marginally do) on the number of publications ($pub$).[39] Formally, we have that $\frac{dTotal\ Cost_{scpub}}{df(.)} * \frac{dK_{res}}{dpub} = \frac{dTotal\ Cost_{scpub}}{df(.)} * \frac{dL_{res'}}{dpub} = \frac{dTotal\ Cost_{scpub}}{df(.)} * \frac{dOP_{res'}}{dpub} = \frac{dTotal\ Cost_{scpub}}{d\gamma(.)} * \frac{dK_{pub}}{dpub} = \frac{dTotal\ Cost_{scpub}}{d\gamma(.)} * \frac{dL_{pub'}}{dpub} = \frac{dTotal\ Cost_{scpub}}{d\gamma(.)} * \frac{dOP_{pub}}{dpub} = 0.$

$$\frac{dTotal\ Cost_{scpub}}{dpub} = \frac{df(L_{res})}{dpub} * \frac{dL_{res}}{dpub} + \frac{d\gamma(L_{pub})}{dpub} * \frac{dL_{pub}}{dpub} > 0 \qquad \text{A.3}$$

Eq. A.3 shows that the total cost of producing scientific publications is an increasing function of the labour cost of scientists ($\frac{dL_{res}}{dpub} > 0$) and publishers ($\frac{dL_{pub}}{dpub} > 0$), namely editors and reviewers. The first term of A.3 can be defined as the marginal production of research ($\frac{df(L_{res})}{dpub} * \frac{dL_{res}}{dpub} \equiv MPC_{res}$) and the second term as the marginal production of publishing ($\frac{d\gamma(L_{pub})}{dpub} * \frac{dL_{pub}}{dpub} \equiv MPC_{pub}$). Assuming the quadratic form for the cost function (A.1), then Eq. A.3 is a linear function of the labour cost, and can be written as follows (Eq. A.4):

$$MPC = MPC_{res} + MPC_{pub} \qquad \text{A.4}$$

which can be further expanded considering the fundamental components entering the marginal production costs and obtaining Eq. 2 in Section 2 in the main text.

$$MPC = MPC_{res} + MPC_{pub} = \left(\frac{w_{res} * h_{res}}{y_{res}}\right) + \left(\frac{w_{pub} * h_{pub}}{y_{pub}}\right) \qquad (A.5)$$

Previous works overlooked the marginal of publishing (Florio, 2019; Florio et al., 2016; Battistoni et al., 2016; EC, 2014). Indeed, setting $Cost_{pub} = 0$ yields Eq. 1 in Section 2 in the main text.

---

[39] A small portion of operational expenses can fluctuate with the number of publications. However, it is assumed that most operational costs—whether for researchers or publishers—remain constant regardless of the number of publications produced, with the consequence that their variation is set to zero.

# Appendix B. The marginal cost of publishing

Following Morretta et al. (2022), the marginal cost of publishing in Eq. 3 can be expressed as:

$$MPC_{pub_t} = \left(\frac{w_{pub_{e,t}} * h_{pub_{i,e,t}}}{y_{pub_{e,t}}}\right) \tag{B.1}$$

where $w_{pub_{e,t}}$ is the average gross annual wage of editors and reviewers ($e$) at the time $t$ employed in the publishing process, $h_{pub_{e,t}}$ is the share of the time they employ in that process, and $y_{pub_{e,t}}$ is the yearly total number of scientific products they peer review and publish. We deliberately omitted the subscript $f$ denoting the scientific field of the authors in the term $MPC_{pub_t}$. Indeed, it is possible for instance that publications in economics are published in journals related to other scientific fields (e.g. physics) and therefore we preferred not to associate $MPC_{pub_t}$ to same scientific fields as the one of scientific products. That said, putting a subscript $d$ that denotes the scientific field of the journal, with $d = f$ or $d \neq f$ does not change the meaning of the formula.

# Appendix C. The paradox of the apportionment of author's research time among the scientific products in her catalogue

Authors within a given scientific field often produce a diverse range of scientific outputs—such as conference proceedings, peer-reviewed articles, notes, and pre-prints—each potentially covering distinct topics. For example, climate science research can encompass areas such as climate system dynamics and modelling, cryosphere studies, or greenhouse gas emissions and carbon cycling. In health sciences, topics might include epidemiology and infectious diseases, geriatrics, chronic diseases, among others. In our context, an author's catalogue of scientific outputs may include both those related to LHC research and other topics common in HEP. **Figure C.1** illustrates a stylised example of an author's catalogue composition in our sample (Set A in Figure C.1). Within her catalogue, an author may have a subset of outputs focused on LHC-related topics (B ⊆ A), and, within this, a subset specifically comprising research articles and pre-prints (C ⊆ B ⊆ A). Beyond LHC-focused articles and pre-prints, the author might also have co-authored other research articles covering distinct HEP topics, forming another subset (D ⊆ A), where C ⊆ B∩D.

**Figure C.1. Representation of a possible author's catalogue of scientific products in HEP**

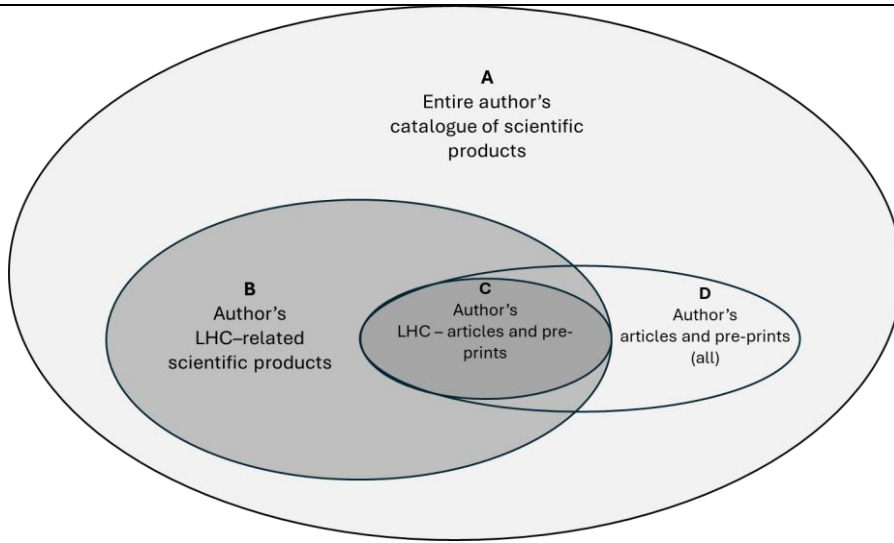

To assign monetary value to specific subsets of an author's publications, whether by topic (e.g., LHC-related research in subset B), type (e.g., research articles in subset C∪D), or a combination thereof (e.g., subset C)— one might initially assume that authors distribute their research time equally across all outputs in their catalogue. However, we demonstrate that this equal-time apportionment yields illogical conclusions, suggesting that outputs in subset B consistently have lower monetary value than those in A, those in C lower than those in B, and so forth, even when the outputs in A, B, and C are identical. This is evidently contradictory. Instead, we argue that a productivity function, using the scale of co-authorship as the principal variable (albeit imperfect), inherently captures the relevant information needed to assign value across the range of scientific outputs of interest. This approach avoids the inconsistencies associated with time-based apportionment and better reflects the value of outputs within each subset.

Starting from Eq.3 in the main text, let us assume we wish to assign a monetary value to a specific scientific product within one of the subsets B, C, or D. For clarity, let's consider the LHC-related products in subset B and let's index it with the subscript $i$. So, we have:

$$MPC_{i,f,t} = MPC^c_{res_{i,f,t}} + MPC_{pub_t} = \left( \frac{w_{res_{j,f,t}} * (h_{res_{j,f,t}} * \alpha_i)}{y_{res_{j,f,t}}(n_t)} \right) * n_{i,t} + MPC_{pub_t} \quad (C.1)$$

The new parameter $\alpha_i$, where $\alpha_i \in (0,1)$ represents a reduction factor accounting for the fact that only a portion of the total research time ($h_{res,j,f,t}$) that author(s) allocate to research is devoted to producing scientific outputs in the subset B. The parameter $\alpha_i$ reaches its maximum value of 1 in the scenario where the author(s) solely produce that $i-th$ product. Conversely, a value of zero would imply no time dedicated to research, which is not relevant for our purposes. Typically, $\alpha_i$ is expected to vary between these two extremes.[40] In the absence of targeted studies, we cannot determine the precise amount of research time the author(s) allocate specifically to the $i-th$ product. Therefore, it may be reasonable to assume that author(s) distribute their time equally across all scientific outputs in their catalogue. Under this assumption, $\alpha_i$ can be set by examining the proportion of LHC-related outputs relative to the total scientific output. As shown in **Figure C.1**, from 1990 to 2021, an average of 22% of the total output by authors in our sample was associated with LHC-related topics.[41]

---

[40] To note that the author(s) salary ($w_{res_{j,f,t}}$) and productivity ($y_{res_{j,f,t}}(n_t)$) are not indexed to $i$ because they do not vary according to the type of scientific product.

[41] Figure C.1 shows the share of LHC-related scientific products in relation to total scientific production considering all the authors of the sample. The LHC share ranges between 31-61% in the 2010 – 2021 period when the LHC was in its operating phase indicating a relative shift in their focus toward LHC- related work as compared to previous years.

**Figure C.1 Share of LHC-related scientific products out of authors' total scientific production (1990 – 2021)**

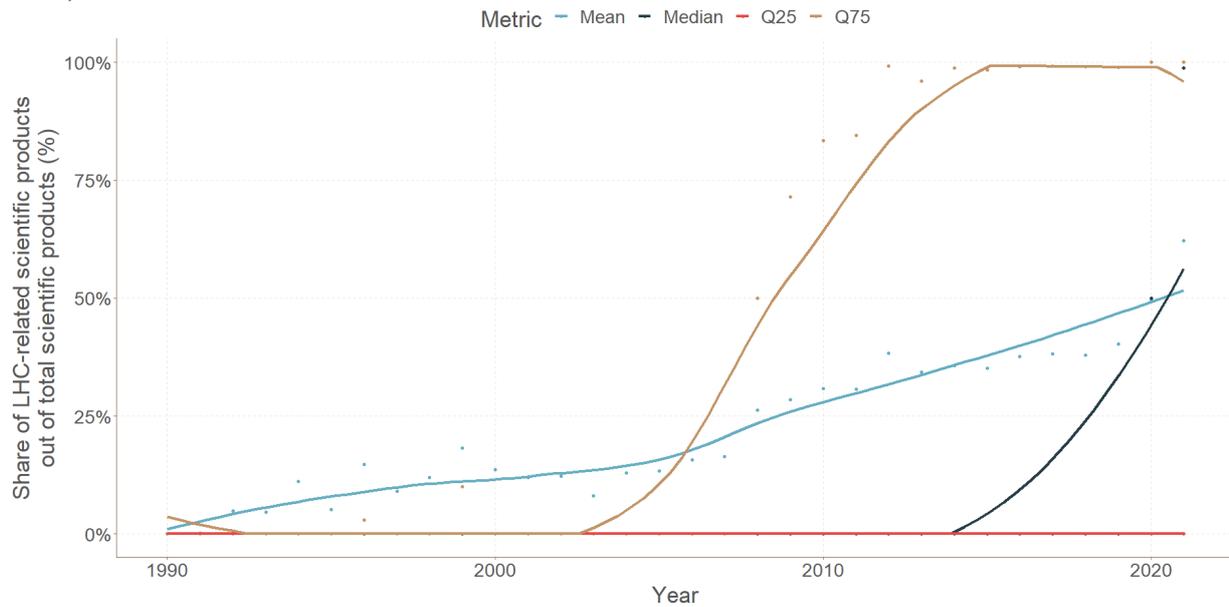

Source: Authors based on INSPIRE HEP data. Full sample of 54,384 retrievable authors with at least one LHC-related scientific product in 1990-2021. Outlying products with more than 3 000 co-authors were excluded. Note: On a given year, all active researchers are considered, even if they did not produce any LHC-related scientific product. For all the relevant authors, we compute the number of LHC-related and total products (including non-LHC-related). The shares of LHC-related products can then be computed for each author on a given year. The distribution of these shares is then described using summary statistics.

Accordingly, the parameter $\alpha_i$ is set at 0.22, resulting in the distribution of monetary values for LHC-related products illustrated in **Figure C.2.** As expected, because of the application of the reduction factor $\alpha_i$, the distribution is right skewed compared to the distribution of all scientific products in set A (**Figure 11** in Section 6.1.1), with lower values. The monetary value of LHC-related outputs now spans from a minimum of EUR PPS 2,460 for single-author contributions to a maximum of EUR PPS 344,408 for extensively collaborative LHC products, involving up to 3,000 co-authors. The median and the mean values are EUR PPS 2,460 and EUR PPS 31,736, respectively (instead, they were EUR PPS 38,222 and EUR PPS 84,210 for products in Set A).

**Figure C.2. Distribution of the monetary values of LHC-related scientific products by applying the reduction factor $\alpha$ (1990 - 2021).**

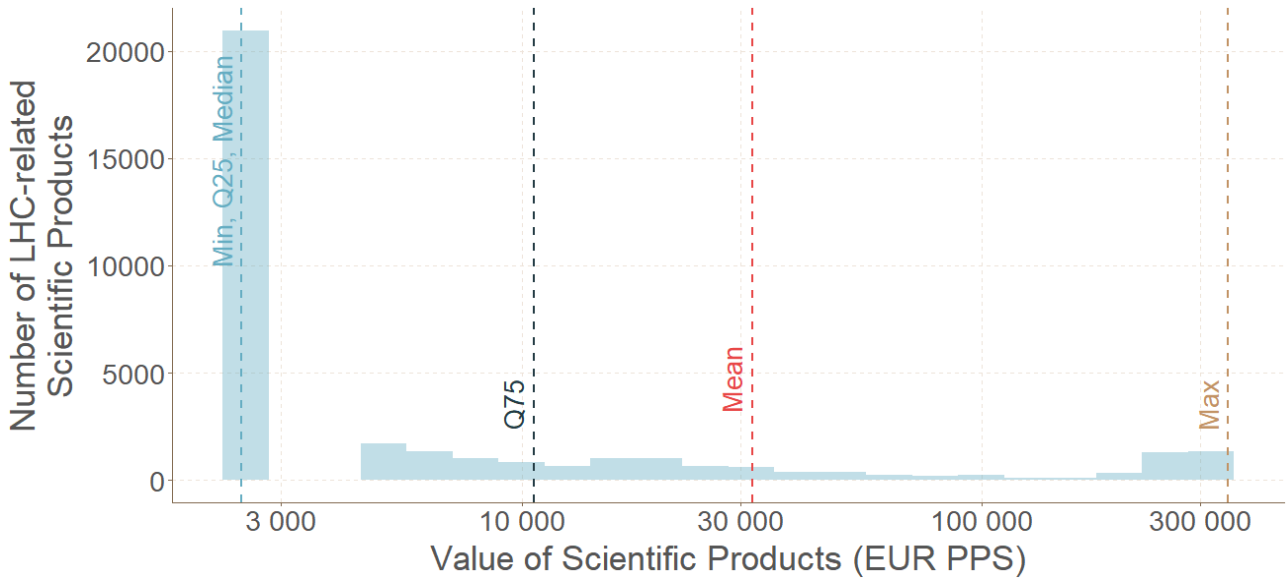

Source: Authors based on Eq. C.1 applied to the empirical distribution of LHC-related scientific products. The sample consists of 34,397 LHC-related scientific products trimmed at 3,000 co-authors (1990-2021). The value of the parameters entering Eq. C.1 were set as at their average values: $\overline{MPC}_{pub} = 311$; $\bar{h}_{res} = 65\%$, $\overline{w}_{res} = 47,837$, $\bar{\alpha} = 0.22$.

Let's proceed and suppose now suppose we wish to calculate the monetary value of research articles and pre-prints within subset C⊆ B. Applying the assumption of equal time allocation requires us to further adjust the reduction factor $\alpha_i$ by the proportion of research articles and pre-prints in the subset C relative to the total LHC-related products in B. In our sample, this proportion is 16.3%. Let's denote it with $\gamma = 0.163$, and apply as a multiplicative factor for $\alpha_i$ in Eq. C.1, i.e., $w_{res_{j,f,t}} * (h_{res_{j,f,t}} * \alpha_i * \gamma)$, with all other parameters remaining constant. The resulting distribution is shown in **Figure C.3**, where now the monetary value of an LHC research article or pre-print ranges from a minimum of EUR PPS 661 to a maximum of EUR PPS 56,399, with a mean of EUR PPS 26,997, namely, 1.18 times lower than the mean value of products in the set B, and 3.13 times lower than the mean value of products in the set A. Due to the left-skewed nature of this distribution, the median is EUR PPS 34,973, 14.2 times higher than products in set B, but 0.91 times lower than products in set A. This introduces a paradox: the same product, such as an LHC research article with 3,000 co-authors belonging to the set C⊆ B⊆ A is progressively undervalued at each level, revealing a clear contradiction. A simplified numerical example is provided in table C.1 below.

**Figure C.3. Distribution of the monetary values of LHC-related articles and pre-prints by applying the reduction factor $\alpha$ and an additional reduction factor $\gamma$ (1990 - 2021).**

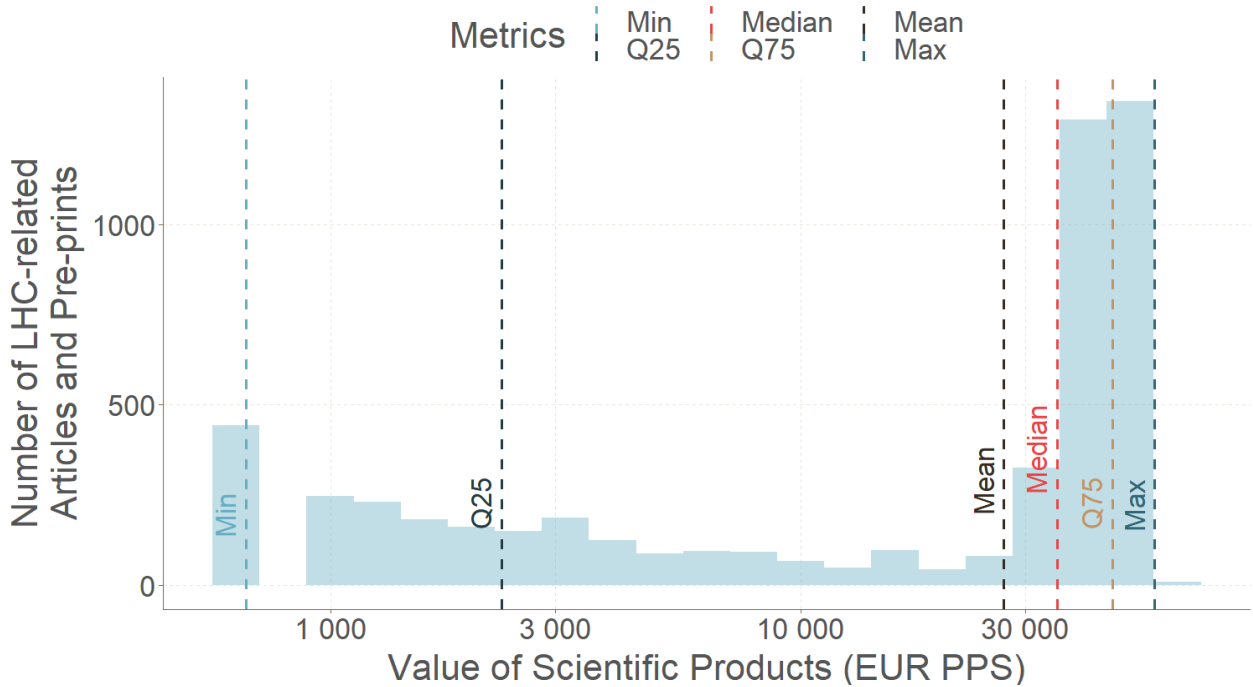

Source: Authors based on Eq. C.1 applied to the empirical distribution of LHC articles and pre-prints. The sample consists of 5,292 products with maximum 3,000 co-authors (1990-2021). The value of the parameters entering Eq. C.1 were set as at their average values: $\overline{MPC}_{pub} = 311$; $\bar{h}_{res} = 65\%$, $\overline{w}_{res} = 47,837$, $\bar{\alpha} = 0.22$ and $\bar{\gamma} = 0.163$

**Table C.1. The paradox of the equal apportionment of research time among the scientific products in the author(s)' catalogue: numerical examples**

| |
|---|
| **PANEL A: Example 1** |
| Suppose an author has two scientific products in a year, one LHC-related indexed by $i$ and the other non-LHC related (indexed by $j$). So, her catalogue only consists of $i$ and $j$ with A = ($i \cup j$), and with $i$ also belonging to the sub-set B ($i \in B$). For the sake of simplicity, let's suppose that she is the only author of both products, so we have n = 1, that her annual salary ($\overline{w}_{res}$) is 100, and the time she dedicates to research ($h_{res}$) equals to 65%. Her productivity function ($y(n)$) is 2 by definition. |
| The evaluation of the two products by applying the marginal cost formula to the entire author's catalogue (no reduction factor is needed) would yield a mean monetary value per product of EUR PPS 32.5 = $\frac{100*0.65}{2}$. Put differently, each product values EUR PPS 32.5 |
| Let's now focus on the LHC-related product ($i$). The most neutral assumption that one can do is to assume that 50% of the author's research time has been dedicated to producing it. Accordingly, $\alpha$ is set at 0.5. The Eq. C.1 returns a monetary value of the product $i$ equal to EUR PPS 16.25 = $\frac{100*0.65*0.5}{2}$, halved compared to the value of the same product when considering the entire catalogue, which was to be demonstrated. |
| **PANEL B: Example 2** |
| Let's consider an author in the health scientific field who produces 3 scientific products in a year, 2 dealing with infectious diseases and 1 covering cancer treatment-related topics. Let's suppose that the infectious diseases products consist of 1 report and 1 peer-reviewed article, and the one related to cancer-treatment is a peer-reviewed article as well. |
| Let's suppose again that her salary annual salary ($\overline{w}_{res}$) is 100, and the time she dedicates to research ($h_{res}$) equals to 65%, and she is the only author of all three products. Her productivity function is now 3 by definition. Considering het total catalogue, we have that the mean value of each product is EUR PPS 21.6 = $\frac{100*0.65}{3}$. |

> Let's suppose that one wants to attribute a monetary value to the peer-reviewed article related to cancer treatment and introduce a reduction factor ($\alpha_i = 0.33$) to reduce her research time dedicated to that article by one-third. Eq. C.1. would return a value of EUR PPS 7.15 = $= \frac{100*0.65*0.33}{3}$, which is again a contradiction.
>
> Similarly, if one wants to attribute a value to the two peer-reviewed articles in her catalogue, then her time should be reduced to two-thirds ($\alpha_i = 0.66$) yielding an average value of EUR PPS 14.3 per article, which differs from 7.15, which in turn differs from 21.6, which was to be demonstrated.

Overall, we argue that the parameter α does not adequately capture the complexity and nuances of the scientific production processes underlying various scientific products. Large international collaborations, for example, operate in a fundamentally different manner from smaller or non-collaborative research efforts, reflecting distinct workflows and structures. Furthermore, scientific outputs themselves are highly heterogeneous. We contend that α-type parameters may be unnecessary altogether, as the productivity function already incorporates the necessary information to evaluate these outputs, although imperfectly.

On top of that, α-like parameters fail to capture quality. To illustrate this with an analogy: if a factory produces three cars—one of low quality, one of medium quality, and one of high quality made with premium materials—would it be logical to assert that the high-quality car represents only one-third of the factory's total production value? Or would it naturally have a higher value than the other two? Our analysis demonstrates that LHC-related products, particularly articles and pre-prints, receive significantly higher citation counts than other type of publications, indicating a greater impact and value within the HEP community. We explore this point further in Section 7.

# Appendix D. Data extraction process

The databases used for the analysis were the result of a series of successive steps based on the INSPIRE HEP API.[42] The following table outlines the main steps of the data extraction and the resulting content.

**Table B.1.** Data extraction process

| Step | Description | Content |
|---|---|---|
| Identification of the LHC collaborations and experiments | Request to the INSPIRE HEP API regarding experiments and collaborations linked to the LHC | 19 LHC collaborations and 31 LHC experiments identified |
| Collection of LHC-related scientific products | • Request to the INSPIRE HEP API to extract all the scientific products related to the LHC collaborations and experiments identified in the previous step.<br>• Collection of variables describing these products<br>• Cleaning the database for duplicates and filtering to restrict it to the 1990-2021 period | • 38,726 LHC-related scientific products (1990-2021) |
| Extraction of authors of LHC-related scientific products | • Extraction of the identifiers of authors from the LHC-related scientific products<br>• Robustness checks to compare the characteristics of products with missing author information from the others | • 35,159 LHC-related scientific products with identified authors (missing information rate 9.2%)<br>• 54 392 distinct authors identified |
| Collection of all scientific products of the extracted authors | • Collection of all the scientific products of the identified authors (LHC-related and non-LHC-related) for the 1990-2021 period<br>• Collection of variables describing these products | 434,065 scientific products retrieved for the 54,384 LHC-related authors of the sample (others were not retrievable). |
| Removal of outliers | Removal of scientific products with more than 3,000 authors | 433,903 scientific products (including 38,564 LHC-related ones, of which 34,997 had information on the names of authors) |

Source: Authors

---

[42] See documentation: https://github.com/inspirehep/rest-api-doc

# Appendix E. Additional statistics

**Table E.1 4. Average gross salary in research in scientists' country in the year 2021 (baseline year: 2006)**

| Country | Avg. gross salary | |
| --- | --- | --- |
| | EUR | PPS |
| Australia | 93,128 | 88,821 |
| Austria | 81,125 | 77,369 |
| Belgium | 73,856 | 69,675 |
| Brazil | 15,079 | 34,230 |
| Canada | 94,660 | 90, 556 |
| China | 5,370 | 22,655 |
| Czechia | 24,890 | 45,708 |
| Finland | 58,116 | 46,876 |
| France | 59,976 | 55,523 |
| Germany | 68,768 | 64,396 |
| Greece | 29,062 | 34,698 |
| India | 21,997 | 93,636 |
| Italy | 44,683 | 41,740 |
| Israel | 57,490 | 79,579 |
| Japan | 65,383 | 58,908 |
| South Korea | 60,223 | 51,854 |
| Mexico | 32,700 | 28,000 |
| Netherlands | 71,013 | 66,678 |
| Norway | 90,553 | 60,670 |
| Poland | 15,847 | 29,005 |
| Portugal | 36,048 | 40,787 |
| Russia | 12,800 | 20,800 |
| Spain | 40,594 | 44,716 |
| Sweden | 73,437 | 60,318 |
| Switzerland | 86,546 | 62,525 |
| Turkey | 70,657 | 98,031 |
| United Kingdom | 74,419 | 68,608 |
| United States | 78,178 | 79,723 |
| Worldwide (weighted average) | 60,553 | 60,656 |

Source: Authors' elaborations based on Morretta et al. (2022). New elaborations refer to years from 1990 to 1997 and from 2019 to 2021 and to the following countries: Brazil, Canada, South Korea del Sud, Mexico, Russia. Value in PPS have been obtained by using the World Bank's global PPP conversion factor. Weights entering the worldwide weighted average are those in Figure 10 in the main text,

**Table E.2.** Average gross salary in research by year (baseline year: 2006)

| Year | Avg. gross salary | |
|---|---|---|
| | EUR | PPS |
| 1990 | 26,380 | 26,469 |
| 1991 | 26,868 | 26,999 |
| 1992 | 27,369 | 27,544 |
| 1993 | 27,883 | 28,107 |
| 1994 | 28,413 | 28,686 |
| 1995 | 28,957 | 29,283 |
| 1996 | 29,517 | 29,899 |
| 1997 | 30,094 | 31,192 |
| 1998 | 31,185 | 30,952 |
| 1999 | 31,646 | 31,428 |
| 2000 | 32,291 | 32,095 |
| 2001 | 33,167 | 32,997 |
| 2002 | 34,037 | 33,899 |
| 2003 | 34,905 | 34,802 |
| 2004 | 35,766 | 35,734 |
| 2005 | 36,554 | 36,568 |
| 2006 | 37,475 | 37,537 |
| 2007 | 38,457 | 38,623 |
| 2008 | 39,530 | 39,795 |
| 2009 | 40,143 | 40,438 |
| 2010 | 40,549 | 40,962 |
| 2011 | 41,331 | 41,879 |
| 2012 | 42,034 | 42,664 |
| 2013 | 42,599 | 43,300 |
| 2014 | 43,121 | 43,862 |
| 2015 | 43,589 | 44,357 |
| 2016 | 44,050 | 44,880 |
| 2017 | 44,588 | 45,535 |
| 2018 | 45,305 | 46,380 |
| 2019 | 48,235 | 51,713 |
| 2020 | 49,537 | 53,315 |
| 2021 | 50,916 | 55,025 |
| Average (not weighted) | 37,078 | 37,716 |

Source: Authors' elaborations based on Morretta et al. (2022). New elaborations refer to years from 1990 to 1997 and from 2019 to 2021 and to the following countries: Brazil, Canada, South Korea, Mexico, Russia. Value in PPS have been obtained by using the World Bank's global PPP conversion factor.

**Table E.3.** Impact on the average value of scientific products based on different scenarios (holding other variables constant)

| Variables | Decrease compared to baseline value | | |
|---|---|---|---|
|  | **-10%** | **-5%** | **-1%** |
| $n$ (n of co-authors) | -6.6% | -2.5% | -0.2% |
| $W_{res}$ (salary) | -10% | -5% | -1% |
| $h_{res}$ (time dedicated to research) | -15.3% | -7.7% | -1.5% |
| $MCP_{pub}$ (marginal cost of publishing) | -0.04% | -0.02% | -0% |

Source: Authors simulation based on the mentioned deviations from the standard parameters. For the number of authors, the entire distribution is shifted uniformly (e.g., publications with 10 authors increase to 11 authors if the number is scaled by 10%, publications with 100 authors to 110). The number of authors was rounded. Parameters' subscripts omitted for simplicity.

**Figure E.1.** Simulation of the monetary value of a scientific product under baseline assumptions (1990-2021)

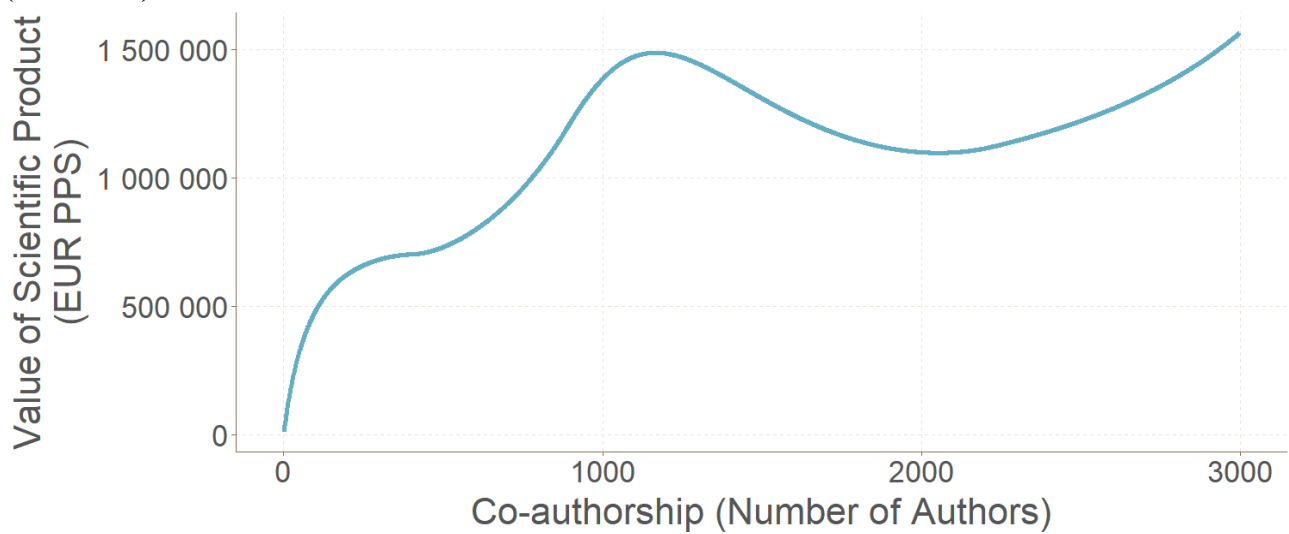

Source: Authors based on equation (3). Apart from the productivity function, the value of the other parameters entering Eq. 3 were set as at their average values: $\overline{MPC}_{pub} = 311; \bar{h}_{res}= 65\%, \overline{w}_{res} = 47{,}837$.

European Commission (2007). Remuneration of Researchers in the public and private sector, April 2007 https://cdn1.euraxess.org/sites/default/files/policy_library/final_report.pdf

European Commission (2014). Guide to Cost Benefit Analysis of Investment Projects. DG Regio, http://ec.europa.eu/regional_policy/sources/docgener/studies/pdf/cba_guide.pdf.

Fan Qi, Hongyu Zhou, Beibei Sun, Ying Huang, & Lin Zhang (2023) "How do boundary-crossing researchers contribute to the interdisciplinary knowledge flows? Evidence from physics." Proceedings of ISSI 2023: 19th International Conference of the International Society for Scientometrics and Informetrics, July 2-5, 2023, Bloomington, IN, USA. Vol. 1. 2023. https://doi.org/10.5281/ZENODO.8305942

FCCIS (2024). Report on Socio-Economic Impacts of the Lepton Collider-Based Research Infrastructure. Future Circular Collider Innovation Study. Horizon 2020 Research and Innovation Framework Programme, Research and Innovation Action. https://doi.org/10.5281/zenodo.10653396

Florio, M. 2019. Investing in Science: Social Cost-Benefit Analysis of Research Infrastructures. Cambridge, MA, United States: MIT Press.

Flores-Szwagrzak, K. &Treibich, R. (2015), Co-authorship and the Measurement of Individual Productivity. University of Southern Denmark Discussion Paper on Business and Economics 17 http://dx.doi.org/10.2139/ssrn.2741594

Florio, M., & Sirtori, E. (2016). Social benefits and costs of large scale research infrastructures. *Technological Forecasting and Social Change*, 112, 65-78. https://doi.org/10.1016/j.techfore.2015.11.024

Florio, M., Forte, S., & Sirtori, E. (2016). Forecasting the socio-economic impact of the Large Hadron Collider: A cost–benefit analysis to 2025 and beyond. *Technological Forecasting and Social Change*, 112, 38-53. https://doi.org/10.1016/j.techfore.2016.03.007

Furnham, A. (2021). Publish or perish: rejection, scientometrics and academic success. *Scientometrics*, 126(1), 843-847. https://doi.org/10.1007/s11192-020-03694-0

Gentil-Beccot, A., Mele, S., & Brooks, T. (2010). Citing and reading behaviours in high-energy physics. *Scientometrics*, 84(2), 345-355. https://doi.org/10.1007/s11192-009-0111-1

Giffoni, F., & Florio, M. (2023). Public support of science: A contingent valuation study of citizens' attitudes about CERN with and without information about implicit taxes. *Research Policy*, 52(1), 104627.

Gómez, L. D. (2015). Does Co-authorship Lead to Higher Academic Productivity? 385-407. Oxford bulletin of economics and statistics, 77(3), 385-407. https://doi.org/10.1016/j.respol.2022.104627

Goodwin, C. D., & Deneef, A. L. (Eds.). (2020). The Academic's Handbook. Duke University Press.

Ioannidis, J. P., Collins, T. A., & Baas, J. (2023). Evolving patterns of extremely productive publishing behavior across science. *bioRxiv*, 2023-11. https://doi.org/10.1101/2023.11.23.568476

Isfandyari-Moghaddam, A., Saberi, M. K., Tahmasebi-Limoni, S., Mohammadian, S., & Naderbeigi, F. (2023). Global scientific collaboration: A social network analysis and data mining of the co-authorship networks. *Journal of Information Science*, 49(4), 1126-1141. https://doi.org/10.1177/01655515211040654